\documentclass[nofootinbib,aps,superscriptaddress,preprint,floatfix]{revtex4}

\usepackage{graphicx,epsf}

\usepackage{bm}

\def\D0bar{\overline D{}^0}
\def\K0bar{\overline K{}^0}

\def\BsBsbar{B{}^0_s-\overline B{}^0_s}

\def\beq{\begin{equation}}
\def\eeq{\end{equation}}
\def\beqa{\begin{eqnarray}}
\def\eeqa{\end{eqnarray}}
\def\bea{\begin{eqnarray}}
\def\eea{\end{eqnarray}}
\def\beq{\begin{equation}}
\def\eeq{\end{equation}}

\def\Re{{\cal R \mskip-4mu \lower.1ex \hbox{\it e}\,}}
\def\Im{{\cal I \mskip-5mu \lower.1ex \hbox{\it m}\,}}
\def\be{\begin{equation}}
\def\ee{\end{equation}}

\def\Re{{\cal R \mskip-4mu \lower.1ex \hbox{\it e}\,}}
\def\Im{{\cal I \mskip-5mu \lower.1ex \hbox{\it m}\,}}

\def\sub#1{_{\lower.25ex\hbox{$\scriptstyle#1$}}}
\def\sul#1{_{\kern-.1em#1}}
\def\sll#1{_{\kern-.2em#1}}
\def\sbl#1{_{\kern-.1em\lower.25ex\hbox{$\scriptstyle#1$}}}
\def\ssb#1{_{\lower.25ex\hbox{$\scriptscriptstyle#1$}}}
\def\sbb#1{_{\lower.4ex\hbox{$\scriptstyle#1$}}}

\def\to{\rightarrow}
\def\dmix{\ifmmode D^0-\bar D^0 \else $D^0$-$\bar D^0$\fi}
\def\dm{\Delta M_D}

\def\dmd{\ifmmode \Delta M_D \else $\Delta M_D$\fi}
\def\mh{\ifmmode m\sbl H \else $m\sbl H$\fi}
\def\mch{\ifmmode M_{H^\pm} \else $M_{H^\pm}$\fi}
\def\mt{\ifmmode m_t\else $m_t$\fi}
\def\mc{\ifmmode m_c\else $m_c$\fi}
\def\mz{\ifmmode M_Z\else $M_Z$\fi}
\def\mw{\ifmmode M_W\else $M_W$\fi}
\def\mws{\ifmmode M_W^2 \else $M_W^2$\fi}
\def\mhs{\ifmmode M_H^2 \else $M_H^2$\fi}
\def\mzs{\ifmmode M_Z^2 \else $M_Z^2$\fi}
\def\mts{\ifmmode m_t^2 \else $m_t^2$\fi}
\def\mcs{\ifmmode m_c^2 \else $m_c^2$\fi}
\def\mchs{\ifmmode M_{H^\pm}^2 \else $M_{H^\pm}^2$\fi}
\def\ztwo{\ifmmode Z_2\else $Z_2$\fi}
\def\zone{\ifmmode Z_1\else $Z_1$\fi}
\def\mtwo{\ifmmode M_2\else $M_2$\fi}
\def\mone{\ifmmode M_1\else $M_1$\fi}
\def\tb{\ifmmode \tan\beta \else $\tan\beta$\fi}
\def\xw{\ifmmode x\sub w\else $x\sub w$\fi}
\def\ch{\ifmmode H^\pm \else $H^\pm$\fi}
\def\lum{\ifmmode {\cal L}\else ${\cal L}$\fi}
\def\inpb{\ifmmode {\rm pb}^{-1}\else ${\rm pb}^{-1}$\fi}
\def\infb{\ifmmode {\rm fb}^{-1}\else ${\rm fb}^{-1}$\fi}
\def\epem{\ifmmode e^+e^-\else $e^+e^-$\fi}
\def\ppb{\ifmmode \bar pp\else $\bar pp$\fi}

\newskip\zatskip \zatskip=0pt plus0pt minus0pt
\def\matth{\mathsurround=0pt}

\def\atversim#1#2{\lower0.7ex\vbox{\baselineskip\zatskip\lineskip\zatskip
  \lineskiplimit 0pt\ialign{$\matth#1\hfil##\hfil$\crcr#2\crcr\sim\crcr}}}

\def\Re{{\cal R \mskip-4mu \lower.1ex \hbox{\it e}\,}}
\def\Im{{\cal I \mskip-5mu \lower.1ex \hbox{\it m}\,}}

\def\sub#1{_{\lower.25ex\hbox{$\scriptstyle#1$}}}
\def\sul#1{_{\kern-.1em#1}}
\def\sll#1{_{\kern-.2em#1}}
\def\sbl#1{_{\kern-.1em\lower.25ex\hbox{$\scriptstyle#1$}}}
\def\ssb#1{_{\lower.25ex\hbox{$\scriptscriptstyle#1$}}}
\def\sbb#1{_{\lower.4ex\hbox{$\scriptstyle#1$}}}

\def\to{\rightarrow}

\def\rb{\ifmmode R_b\else $R_b$\fi}
\def\rc{\ifmmode R_c\else $R_c$\fi}
\def\ac{\ifmmode A_c\else $A_c$\fi}
\def\dmix{\ifmmode D^0-\bar D^0 \else $D^0$-$\bar D^0$\fi}
\def\dm{\ifmmode \Delta M_D \else $\Delta M_D$\fi}
\def\rb{\ifmmode R_b\else $R_b$\fi}
\def\mh{\ifmmode m\sbl H \else $m\sbl H$\fi}
\def\mch{\ifmmode M_{H^\pm} \else $M_{H^\pm}$\fi}
\def\mt{\ifmmode m_t\else $m_t$\fi}
\def\mc{\ifmmode m_c\else $m_c$\fi}
\def\mz{\ifmmode M_Z\else $M_Z$\fi}
\def\mw{\ifmmode M_W\else $M_W$\fi}
\def\mws{\ifmmode M_W^2 \else $M_W^2$\fi}

\def\mhs{\ifmmode m_H^2 \else $m_H^2$\fi}
\def\mzs{\ifmmode M_Z^2 \else $M_Z^2$\fi}
\def\mts{\ifmmode m_t^2 \else $m_t^2$\fi}
\def\mcs{\ifmmode m_c^2 \else $m_c^2$\fi}
\def\mchs{\ifmmode m_{H^\pm}^2 \else $m_{H^\pm}^2$\fi}
\def\ztwo{\ifmmode Z_2\else $Z_2$\fi}
\def\zone{\ifmmode Z_1\else $Z_1$\fi}
\def\mtwo{\ifmmode M_2\else $M_2$\fi}
\def\mone{\ifmmode M_1\else $M_1$\fi}
\def\bsg{\ifmmode b\to s\gamma\else $b\to s\gamma$\fi}
\def\tb{\ifmmode \tan\beta \else $\tan\beta$\fi}
\def\xw{\ifmmode x\sub w\else $x\sub w$\fi}
\def\ch{\ifmmode H^\pm \else $H^\pm$\fi}
\def\lum{\ifmmode {\cal L}\else ${\cal L}$\fi}
\def\inpb{\ifmmode {\rm pb}^{-1}\else ${\rm pb}^{-1}$\fi}
\def\infb{\ifmmode {\rm fb}^{-1}\else ${\rm fb}^{-1}$\fi}
\def\epem{\ifmmode e^+e^-\else $e^+e^-$\fi}
\def\ppb{\ifmmode \bar pp\else $\bar pp$\fi}

\def\be{\begin{equation}}
\def\ee{\end{equation}}

\begin{document}
\vspace{3.0cm}
\preprint{\vbox 
{
\hbox{WSU--HEP--1101} \hbox{UH-511-1136-09}
}}

\vspace*{2cm}

\title{\boldmath 
Relating $B_{s}$ Mixing and $B_{s} \to \mu^+\mu^-$ 
with New Physics}

\author{Eugene Golowich}
\affiliation{Department of Physics,
        University of Massachusetts\\[-6pt]
        Amherst, MA 01003}

\author{JoAnne Hewett}
\affiliation{SLAC National Accelerator Laboratory, 2575 Sand Hill Rd, 
Menlo Park, CA, 94025, USA}

\author{Sandip Pakvasa}
\affiliation{Department of Physics and Astronomy\\[-6pt]
        University of Hawaii, 
        Honolulu, HI 96822}

\author{Alexey A.\ Petrov}
\affiliation{Department of Physics and Astronomy\\[-6pt]
        Wayne State University, Detroit, MI 48201}

\affiliation{Michigan Center for Theoretical Physics\\[-6pt]
        University of Michigan, Ann Arbor, MI 48196\\[-6pt] $\phantom{}$ }

\author{Gagik K.\ Yeghiyan\vspace{8pt}}
\affiliation{Department of Physics and Astronomy\\[-6pt]
        Wayne State University, Detroit, MI 48201}

\begin{abstract}

We perform a study of the Standard Model (SM) fit to 
the mixing quantities $\Delta M_{B_s}$, and 
$\Delta \Gamma_{B_s}/\Delta M_{B_s}$ in order to bound 
contributions of New Physics to $B_s$ mixing.  
We then use this to explore the branching fraction of 
$B_{s} \to \mu^+\mu^-$ in certain models of New Physics (NP).  
In most cases, this constrains NP amplitudes for 
$B_{s} \to \mu^+\mu^-$ to lie below the SM component. 
\vskip 1in
\end{abstract}

\def\thepage{{}}
\maketitle
\def\thepage{\arabic{page}}

\section{Introduction}

We report here on a study of New Physics (NP) 
predictions for $B_s \to \mu^+ \mu^-$.  
The Standard Model (SM) prediction for $B_s \to \mu^+\mu^-$ 
is currently smaller than the experimental branching fraction 
limit~\cite{PDG} of ${\cal B}_{B_s \to \mu^+\mu^-}^{\rm (expt)}$ 
by about a factor of $15$.  This presents a window of opportunity 
for observing New Physics (NP) effects in this mode.  

This topic is particularly timely in view of 
experimental indications of NP effects in both 
the exclusive decay  $B_s\to J/\Psi + \Phi$~\cite{Bona:2008jn} 
(for recent CDF results, also see Ref.~\cite{Louise}) 
as well as 
the inclusive like-sign dimuon asymmetry observed in 
$p{\bar p} \to \mu\mu + X$~\cite{Abazov:2010hv}.  Moreover, 
future work at LHC-B, $e^+e^-$ Super B-factories and 
ongoing CDF $\&$ D0 measurements at Fermilab (see the 
discussion following Eq.~(\ref{bmm}))
is expected to markedly improve the current branching fraction bound. 


Our strategy in this paper is somewhat reminiscent of our recent 
study~\cite{Golowich:2009ii} noting that in some NP models the 
$D^0$ mixing and $D^0 \to \mu^+ \mu^-$ decay amplitudes 
have a common dependence on the NP parameters.  
If so, one can predict the $D^0 \to \mu^+ \mu^-$ branching 
fraction in terms of the observed $\Delta M_{D}$ 
provided that much or all of the mixing is attributed 
to NP.  This is a viable possibility for 
$D^0$ mixing because the Standard Model (SM) signal has large 
theoretical uncertainties and because many NP models can produce 
the observed mixing~\cite{Golowich:2007ka}.  

For $\Delta M_{B_s}$ the situation is very different.
Here, the SM prediction is in accord with the observed value 
({\it e.g.} see Refs.~\cite{{Buras:2003td},{Lenz:2006hd}} 
and papers cited therein).  In fact, the analysis described 
below ({\it cf.} see Eqs.~(\ref{av-np}),(\ref{av-onesig}))
gives $|\Delta M^{\rm (NP)}_{B_s}/
\Delta M^{\rm (SM)}_{B_s}|\le 0.20$, which demonstrates 
just how well the SM prediction agrees 
with the experimental value of $\Delta M_{B_s}$. 
In view of this, our SM expression for $\Delta M_{B_s}$ will be 
given at NLO~\cite{{Buras:1990fn},{Urban:1997gw}} 
whereas LO results will suffice for NP models. As 
regards the corresponding width difference 
$\Delta \Gamma_{B_s}$, the experimental and theoretical 
uncertainties are still rather significant ({\it viz} Sect.~II-C).

In those NP models where 
mixing and $B_s \to \mu^+\mu^-$ arise from a common set of 
parameters, the severe constraint on any NP signal to $B_s$ mixing 
places strong bounds on its contribution to 
${\cal B}_{B_s \to \mu^+\mu^-}$.\footnote{In particular, 
Ref.~\cite{Buras:2003td} considers the possibility, not covered here, 
on effects of so-called minimal flavor violation which  
affect the quark mixing-matrix elements.}  In fact, we shall find 
the constraint can be so strong that for some NP models 
the predicted $B_s \to \mu^+\mu^-$ branching fraction lies 
well below the SM prediction.

The first step in our study ({\it cf}~ Section II) will be to 
revisit the SM predictions for mixing in the $b$-quark system 
by using up-to-date inputs.  We carry this out for 
the two mixing quantities $\Delta M_{B_s}$ and 
$\Delta \Gamma_{B_s}/\Delta M_{B_s}$.  
The former in turn yields phenomenological bounds on NP mixing 
contributions which in 
certain models can be used to bound the magnitude of 
the $B_s \to \mu^+ \mu^-$ decay mode.  We also update the 
SM branching fraction for $B_s \to \mu^+ \mu^-$ by using the 
observed $B_s$ mixing as input.  
Then, in Section III we discuss general properties of 
NP models with tree-level amplitudes.  
In Section IV, we explore various NP models such as 
extra $Z'$ bosons, family symmetry, 
R-parity violating supersymmetry, flavor-changing Higgs models, 
and models with the fourth sequential generation. 
Our concluding remarks appear in Section V, 
and some technical details are relegated to the Appendix.

\begin{table}[b]
\begin{tabular}{l|l}
\colrule\hline 
\colrule \colrule
$M_{B_s} = 5366.3 \pm 0.6$~MeV \cite{PDG}   & 
$\tau_{B_s} = (1.425 \pm 0.041)\times 10^{-12}$~s \cite{PDG}
\\
$\Delta M_{B_s} = 
(117.0 \pm 0.8) \times 10^{-13}~{\rm GeV}$ & 
$\Delta \Gamma_{B_s}/\Gamma_{B_s} = 0.092_{-0.054}^{+0.051}$
\cite{PDG}   
 \\
$x_{B_d} = 0.776  \pm 0.008$ \cite{PDG} & 
$x_{B_s} = 26.2  \pm 0.5$  \cite{PDG} \\
$m_t^{\rm (pole)} = 173.1 \pm 1.3$  \cite{Datta:2009} & 
$\alpha_s (M_Z) = 0.1184 \pm 0.0007$ \cite{Bethke:2009jm}\\
$f_{B_s} = 0.2388 \pm 0.0095$~GeV \cite{Laiho:2009eu} 
& $f_{B_s}\sqrt{{\hat B}_{B_s}} = 275 \pm 13$~MeV  \cite{Laiho:2009eu}\\
$|V_{ts}| = 0.0403^{+0.0011}_{-0.0007}$ \cite{PDG} & 
$|V_{tb}| = 0.999152^{+0.000030}_{-0.000045}$ \cite{PDG} \\
\colrule\hline
\end{tabular}
\vskip .05in\noindent
\caption{List of Input Parameters}
\label{tab:corr}
\end{table}
\section{Update of $B_s$ Mixing and $B_s \to \mu^+\mu^-$ in the 
Standard Model}
\noindent We begin by considering the SM predictions for $B_s$ mixing. 
This step is crucial to 
obtaining bounds on NP contributions.  We also use the $B_s$ mixing
signal as input to a determination of the branching fraction for 
$B_s \to \mu^+\mu^-$.
\subsection{Inputs to the Analysis}
The work in this Section takes advantage of recent progress made 
in determining several quantities used in the analysis.
We summarize our numerical inputs in Table I, along with 
corresponding references.  Included in Table I is 
an updated determination of the top quark pole 
mass~\cite{Datta:2009}  
$m_t^{\rm (pole)}$ which in turn is used to determine 
the corresponding running mass 
${\bar m}_t ({\bar m}_t)$~\cite{Melnikov:2000qh} along with 
several decay constants and B-factors 
as evaluated in lattice QCD. 
For definiteness, we have used values appearing in 
Ref.~\cite{Laiho:2009eu}.
This area is, however, constantly evolving and one anticipates 
further developments in the near future~\cite{Aida}. 
Our values for the Cabibbo-Kobayashi-Maskawa (CKM) matrix 
elements $|V_{ts}|$ and $|V_{tb}|$ are taken from Ref.~\cite{PDG}.  
Similar values occur for the global fits cited 
elsewhere~({\it e.g.}~Refs.~\cite{{Antonelli:2009ws},{ckmfitter}}).

\subsection{$\Delta M_{B_s}$}

\noindent The PDG value for $\Delta M_{B_s}$, 
\beq
\Delta M_{B_s}^{\rm (expt)} = (117.0 \pm 0.8) \times 10^{-13}~{\rm GeV} 
 \ \ , 
\label{mexpt}
\eeq
is a very accurate one -- the uncertainty amounts to about $0.7$\%.
The NLO SM formula, 
\beq
\Delta M_{B_s}^{\rm (SM)} = 
2 {G_{\rm F}^2 M_{\rm W}^2 M_{B_s} 
 f^2_{B_s}{\hat B}_{B_s} \over 12 \pi^2} |V_{\rm ts}^* V_{\rm tb}|^2 
\eta_{B_s} S_0({\bar x}_t)\ \ ,
\label{delmSM}
\eeq
is arrived at from an operator product expansion of the mixing 
hamiltonian. The short-distance dependence in 
the Wilson coefficient appears in the scale-insensitive 
combination $\eta_{B_s} 
S_0({\bar x}_t)$, where 
the factor $S_0({\bar x}_t)$ is an Inami-Lin
function~\cite{Inami:1980fz} 
(with ${\bar x}_t \equiv {\bar m}^2_t 
({\bar m}_t)/M_{\rm W}^2$) and 
${\bar m}_t({\bar m}_t)$ is the running top-quark mass parameter 
in ${\overline{{\rm MS}}}$ renormalization.  
In particular, we have 
${\bar m}_t ({\bar m}_t) = (163.4 \pm 1.2)~{\rm GeV}$ which leads to  
$S_0({\bar x}_t) = 2.319 \pm 0.028$.  Using the same matching 
scale, we obtain $\eta_{B_s} = 0.5525 \pm 0.0007$ for the NLO QCD factor. 

Our evaluation for $\Delta M_{B_s}^{\rm (SM)}$ then 
gives
\beqa
& & \Delta M_{B_s}^{\rm (SM)} = 
\left( 125.2^{+13.8}_{-12.7}\right) \times 10^{-13}~{\rm GeV} 
 \ \ , 
\label{me}
\eeqa
which is in accord with the experimental value of 
Eq.~(\ref{mexpt}).
The theoretical uncertainty in the SM prediction of Eq.~(\ref{me}) 
is roughly a factor of sixteen larger than the experimental uncertainty of 
Eq.~(\ref{delmSM}).  
The largest source of error occurs in the 
nonperturbative factor ${\hat B}_{B_s} f^2_{B_s}$, followed by that in 
the CKM matrix element $V_{ts}$.  The asymmetry in the upper and 
lower uncertainties in $\Delta M_{B_s}^{\rm (SM)}$ 
arises from the corresponding asymmetry in the value of 
$V_{ts}$ cited in Ref.~\cite{PDG}. 

Finally, we note in passing that for the ratio 
${\Delta M_{B_d}/\Delta M_{B_s}}$ the experimental value is 
$0.02852 \pm 0.00034$ whereas the SM determination gives 
$0.02835 \pm 0.00187$.  This good agreement is not surprising since 
the ratio ${\Delta M_{B_d}/\Delta M_{B_s}}$ contains less 
theoretical uncertainty than 
$\Delta M_{B_d}$ or $\Delta M_{B_s}$ separately.

\subsection{The Ratio $\Delta\Gamma_{B_s}/\Delta M_{B_s}$}

The above discussion of $\Delta M_{B_s}^{\rm (SM)}$ 
sets the stage for analyzing NP contributions to 
$B_s \to \mu^+\mu^-$.  There is, in principle, a second approach 
which instead utilizes $\Delta\Gamma_{B_s}$.  
The PDG value for the $B_s$ width difference is 
$\Delta \Gamma_{B_s}^{\rm (expt)} = 0.062^{+0.034}_{-0.037} \times 
10^{12}{\rm s}^{-1}$.  Together with Eq.~(\ref{mexpt}), 
this gives\footnote{Using instead the recent CDF evaluation 
$\Delta \Gamma_{B_s}^{\rm (CDF)} = 
{0.075 \pm 0.035 \pm 0.01} \times 10^{12}~{\rm s}^{-1}$ 
implies 
$r^{\rm (expt)} = (42.2 \pm 20.5) \times 10^{-4}$, 
consistent with the value in Eq.~(\ref{bsratio}).}
\beq
r^{\rm (expt)} \equiv { \Delta \Gamma_{B_s}^{\rm (expt)} 
\over \Delta M_{B_s}^{\rm (expt)}} = 
{0.062^{+0.034}_{-0.037} \times 10^{12}~{\rm s}^{-1}\over 
(17.77 \pm 0.12) \times 10^{12}~{\rm s}^{-1}}  = 
(34.9 \pm 20.0) \times 10^{-4} \ \ .
\label{bsratio}
\eeq
whereas the corresponding SM prediction from 
Ref.~\cite{Lenz:2006hd} is $r^{\rm (SM)} = (49.7 \pm 9.4) 
\times 10^{-4}$.  In contrast to the mass splitting 
$\Delta M_{B_s}$, the theoretical uncertainty in the ratio 
$\Delta \Gamma_{B_s}/\Delta M_{B_s}$ is much 
smaller than in the current experimental determination.  
Nonetheless, this situation is expected to change once 
LHCb gathers sufficient data.  As such, we would expect 
a highly accurate value of $\Delta \Gamma_{B_s}^{\rm (expt)}$ 
to eventually become available.  We propose that it could be applied 
to the kind of analysis used in this paper as follows.  
We define a kind of mass difference ${\cal D} M_{B_s}$ as
\beq
{\cal D} M_{B_s} \equiv {\Delta M_{B_s}^{\rm (thy)} \over 
\Delta \Gamma_{B_s}^{\rm (thy)}}~\Delta \Gamma_{B_s}^{\rm (expt)} \ \ .
\label{ratio}
\eeq
The point is that if NP contributions are neglected in 
$\Delta B = 1$ transitions, then $\Delta \Gamma_{B_s}^{\rm (thy)}$ 
is purely a SM effect.  In addition, the ratio 
$\Delta M_{B_s}^{\rm (SM)}/\Delta \Gamma_{B_s}^{\rm (SM)}$ 
will be less dependent on hadronic parameters than either factor 
separately. 

This quantity is also important in the scenarios where NP contributes a 
significant CP-violating phase to $\Delta M_{B_s}$. In this 
situation, $\Delta \Gamma_{B_s}^{\rm (expt)} $ will be reduced 
compared to its SM value $\Delta \Gamma_{B_s}^{\rm (SM)}$ by a 
factor of $\cos 2\xi$, where $\xi$ is related to the relative phase between 
the SM and NP contributions to $\Delta M_{B_s}$~\cite{Grossman:1996era}. 

At the very least, the relation in Eq.~(\ref{ratio}) would be of interest to analyze 
the NP issue using both quantities $\Delta M_{B_s}$ and the above 
${\cal D}M_{B_s}$.  

\subsection{$B_s \to \mu^+\mu^-$}

PDG entries for ${\cal B}_{B_s \to \ell^+\ell^-}$ are  
\beq
{\cal B}_{B_s \to \mu^+\mu^-}^{\rm (expt)} 
< 4.7 \times 10^{-8} \qquad \text{and} 
\qquad 
{\cal B}_{B_s \to e^+e^-}^{\rm (expt)} < 2.8 \times 10^{-7} \ \ , 
\label{bmm}
\eeq
with no experimental limit currently 
for the $B_s \to \tau^+\tau^-$ transition.
Data collected by the D0 and CDF collaborations will 
improve the above brancing fraction limit.
For example, the D0 collaboration reports
${\cal B}_{B_s \to \mu^+\mu^-}^{\rm (D0)} 
< 5.1 \times 10^{-8}$, with an anticipated limit of eleven times 
the SM prediction and similarly for the CDF 
collaboration~\cite{casey}. 

Since the long distance (LD) estimate for the branching fraction of 
$B_s \to \mu^+\mu^-$ in the SM gives  
${\cal B}^{\rm (LD)}_{B_s \to \mu^+\mu^-} \sim
6 \times 10^{-11}$~\cite{Burdman:2001tf}, we consider only the 
short distance (SD) component in 
the following.  Using Eq.~(\ref{delmSM}) as input 
to the SD-dominated $B_s \to \mu^+\mu^-$ transition (see also 
Ref.~\cite{Buras:2003td}) 
we arrive at 
\beq
{\cal B}_{B_s \to \mu^+\mu^-}^{\rm (SM)} = 
\Delta M_{B_s} ~ \tau_{B_s} 
{3 G_{\rm F}^2 M_{\rm W}^2 m_\mu^2 \over 4 \eta_{B_s} 
{\hat B}_{B_s} \pi^3} 
\left[ 1 - 4 {m_\mu^2 \over M_{B_s}^2} \right]^{1/2} {\eta_Y^2 
Y^2({\bar x}_t) \over S_0({\bar x}_t)} \ \ , 
\eeq
where $Y({\bar x}_t)$ is another Inami-Lin function~\cite{Inami:1980fz}.  
Expressing ${\cal B}_{B_s \to \mu^+\mu^-}^{\rm (SM)}$ in this manner 
serves to remove some of the inherent model dependence.
Numerical evaluation gives 
\beq
{\cal B}_{B_s \to \mu^+\mu^-}^{\rm (SM)} =
\left(3.33 \pm 0.21 \right) \times 10^{-9} \ \ .
\label{bmmsm}
\eeq
The major sources of uncertainty, ordered by magnitude, 
arise from the factors ${\hat B}_{B_s}$, then 
the lifetime $\tau_{B_s}$ and 
finally the top-quark mass value. 

\section{Study of New Physics Models}
In this section, we first obtain a numerical (1$\sigma$) bound 
on any possible New Physics contribution to $\Delta M_{B_s}$.  We then 
use this to constrain couplings in a variety of NP models 
and thereby learn something about the $B_s \to \mu^+\mu^-$ transition. 

\subsection{Constraints on NP Models from $B_s$ Mixing} 

As shown in Ref.~\cite{Badin:2009ww}, New Physics in 
$\Delta B = 1$ interactions can {\it in principle} markedly 
affect $\Delta \Gamma_s$.  The logic is similar to that used in 
Ref.~\cite{Golowich:2006gq} regarding the possible impact of NP on 
$\Delta \Gamma_D$.  Since, however, 
in $B_s$ mixing such models are not easy to come up
with, one can simply assume that $\Delta B = 1$ processes are 
dominated by the SM interactions.  
Thus we can write 
\beqa\label{DeltaBNP0}
& & \Delta M_{B_s} = \Delta M_{B_s}^{\rm (SM)} + 
\Delta M_{B_s}^{\rm (NP)} \cos\phi\ \ , 
\eeqa
If the $\Delta B = 1$ sector were to contain significant 
NP contributions, then the above relation would no longer be valid 
due to interference between the SM and NP components. 

As can be seen from Eq.~(\ref{DeltaBNP0}), interference between the SM and 
NP components may also occur in the presence of a 
CP-violating phase $\phi$ in the NP part of the mixing amplitude~\cite{Ball:2006xx}.  
This large NP phase could markedly affect 
$\Delta \Gamma_{B_s}^{\rm (expt)}$ even in the absence of a NP contribution to the 
on-shell $\Delta B = 1$ transitions (recall that $\Delta \Gamma_{B_s}^{\rm (expt)}$ depends 
explicitly on the cosine of the CP-violating phase $\xi$~\cite{Lenz:2006hd,Grossman:1996era};
the explicit relation between $\phi$ and $\xi$ can be found in ~\cite{Grossman:1996era}). It is 
therefore more reasonable to use $\Delta \Gamma_{B_s}^{\rm (expt)}$ in studying those scenarios 
with a large NP phase.  The appropriate strategy here would be to use $\Delta \Gamma_{B_s}^{\rm (expt)}$
and $\Delta \Gamma_{B_s}^{\rm (SM)}$ to extract the phase $\xi$, eliminate 
$\cos\phi$ from Eq.~(\ref{DeltaBNP0}), and then extract $\Delta M_{B_s}^{\rm (NP)}$ in order to
relate it to the rare leptonic decay rate. To do so, however, will require 
a significant reduction in the experimental uncertainty of 
$\Delta \Gamma_{B_s}^{\rm (expt)}$. Alternatively, CP-violating phases could be 
extracted at LHCb from the studies of $B_s \to J/\psi \phi$ transition~\cite{Ball:2006xx}. We shall
defer those studies to a future publication~\cite{Future}.
In this paper we shall assume that the phase in the NP component of $\Delta M_{B_s}$ is sufficiently 
small (although not necessarily negligible), 
\beqa\label{DeltaBNP1}
& & \Delta M_{B_s} = \Delta M_{B_s}^{\rm (SM)} + 
\Delta M_{B_s}^{\rm (NP)} \ \ .
\eeqa

\noindent Accounting for NP as an additive contribution, 
\beq
\Delta M_{B_s}^{\rm (expt)} =  \Delta M_{B_s}^{\rm (SM)} + 
\Delta M_{B_s}^{\rm (NP)} \ \ ,   
\eeq
we have from Eqs.~(\ref{mexpt}),(\ref{me}), 
\beqa
& & \Delta M_{B_s}^{\rm (NP)} 
= \left(- 8.2_{-12.7}^{+13.8}\right) \times 10^{-13}~{\rm GeV} \ \ . 
\label{av-np}
\eeqa
The error in $\Delta M_s^{\rm (expt)}$ has been included, but it  
is so small compared to the theoretical error 
in $\Delta M_s^{\rm (SM)}$ as to be negligible.
The $1\sigma$ range for the NP contribution is thus
\beqa
& & \Delta M_{B_s}^{\rm (NP)} = 
(- 20.9 \to + 5.6)\times 10^{-13}~{\rm GeV} \ \ .
\label{av-onesig}
\eeqa
To proceed further without ambiguity, we would need to know the 
relative phase between the SM and NP components.  Lacking this, 
we employ the absolute value of the largest possible number,  
\beqa
& & |\Delta M_{B_s}^{\rm (NP)}| \le 
20.9 \times 10^{-13}~{\rm GeV} \ \ ,
\label{bnd}
\eeqa
to constrain the NP parameters. 

\subsection{Generic NP Models with tree-level amplitudes} 
\label{sec:TreeLevel}
New Physics can affect both $B_s$ mixing 
and rare decays like $B_s \to \mu^+\mu^-$ 
by engaging in these two transitions at tree level.
In this section we will, for generality, consider a 
generic spin-1 boson V or a 
spin-0 boson S with flavor-changing and flavor-conserving neutral 
current interactions that couple both to quarks and leptons. 
The bosons V and S can be of either 
parity. This situation is frequently realized, as in the interactions of a 
heavy $Z^\prime$ boson or 
in multi-Higgs doublet models without natural flavor conservation. 

\vspace{0.3cm}

{\it Spin-1 Boson V}: 
Assuming that the spin-1 particle $V$ has flavor-changing 
couplings, the most general Lagrangian can be written 
as\footnote{Throughout, our convention for 
defining chiral projections for a field $q(x)$ will be 
$q_{L,R}(x) \equiv(1 \pm \gamma_5) q(x)/2$.} 
\bea\label{VecHam}
{\cal H}_V = g_{V1}^\prime \overline \ell_L^\prime \gamma_\mu \ell_L V^\mu + 
g_{V2}^\prime \overline \ell_R^\prime \gamma_\mu \ell_R V^\mu 
+  g_{V1} \overline b_L \gamma_\mu s_L V^\mu + 
g_{V2} \overline b_R \gamma_\mu s_R V^\mu + \mbox{h.c.} \ \ .
\eea
Here $V_\mu$ is the vector field and 
the flavor of the lepton $\ell^\prime$ might or 
might not coincide with $\ell$. It is not important whether the field 
$V_\mu$ corresponds to an abelian or non-abelian gauge symmetry group.
Using methods similar to those in Ref.~\cite{Golowich:2009ii}, we obtain
\bea\label{dMV2}
\Delta M_{{\rm B}_s}^{\rm (V)} 
&=& {f_{B_s}^2 M_{B_s} \over 3 M_{V}^2} ~
{\cal R}e \left[  C_1 (\mu) B_1 + C_6 (\mu) B_6  -
\frac{5}{4} C_2 (\mu) B_2 + \frac{7}{8} C_3 (\mu) B_3  \right] \ \ , 
\eea
where the superscript on $\Delta M_{{\rm B}_s}^{\rm (V)}$ denotes 
propagation of a vector boson in the tree 
amplitude.  The Wilson coefficients evaluated at a scale $\mu$ are 
related to the couplings $g_{V1}$ and $g_{V2}$ as
\bea
\begin{array}{l}
{\rm C}_1(\mu) = r (\mu,M_V)~ g_{V1}^2\ , \nonumber \\
{\rm C}_2(\mu)= 2 \ r(\mu,M_V)^{1/2} g_{V1} g_{V2}   \ , 
\end{array}
\quad 
\begin{array}{l}
{\rm C}_3(\mu)= \frac{4}{3} \left[
r(\mu,M_V)^{1/2} - r(\mu,M_V)^{-4}
\right] g_{V1} g_{V2} \ , \nonumber \\
{\rm C}_6(\mu)= r (\mu,M_V) ~ g_{V2}^2 \ \ , 
\end{array} 
\label{zwilsons}
\eea
where (presuming that $M > m_t$ and $\mu \ge m_b$), 
\beqa
r(\mu,M)&=& \left(\frac{\alpha_s(M)}{\alpha_s(m_t)}\right)^{2/7}
\left(\frac{\alpha_s(m_t)}{\alpha_s(\mu)}\right)^{6/23} \ \ .
\label{wilson}
\eeqa
Similar calculations can be performed for the 
$B_s^0 \to \ell^+\ell^-$ decay.  The effective Hamiltonian 
in this case is 
\beq
{\cal H}_{b\to q\ell^+\ell^-}^{\rm (V)} = \frac{1}{M_V^2} 
\left[
g_{V1}g_{V1}' \widetilde Q_1 +
g_{V1}g_{V2}' \widetilde Q_7 + g_{V1}'g_{V2}  \widetilde Q_2 +
g_{V2}g_{V2}'  \widetilde Q_6 
\right]\ ,
\eeq
where the operators $\{ \widetilde Q_i\}$ can be read off 
from those in Ref.~\cite{Golowich:2009ii} with the label 
changes $c\to s$ and $u \to b$.  This leads to the branching fraction, 
\beq\label{GammaV}
{\cal B}_{B_s^0 \to \ell^+\ell^-}^{\rm (V)} = 
\frac{f_{B_s}^2 m_\ell^2 M_{B_s}}
{32\pi M_V^4 \Gamma_{B_s}}
\sqrt{1-\frac{4 m_\ell^2}{M_{B_s}^2}} ~
|g_{V1} - g_{V2}|^2 |g_{V1}' - g_{V2}'|^2\ .
\eeq
Clearly, Eqs.~(\ref{dMV2}),(\ref{GammaV}) can be related 
to each other only for a specific set of NP models. 

\vspace{0.3cm}

{\it Spin-0 Boson S}: 
Analogous procedures can be followed if now the FCNC is generated 
by quarks interacting with spin-0 particles. Again, the most general Hamiltonian 
can be written as
\bea\label{SFCNC}
{\cal H}_S = g_{S 1}^\prime  \overline \ell_L \ell_R S + 
g_{S 2}^\prime  \overline \ell_R \ell_L S +
g_{S 1} \overline b_L s_R S + 
g_{S 2} \overline b_R s_L S+ \mbox{h.c.} \ \ .
\eea
Evaluation of $\Delta M_{{\rm B}_s}^{\rm (S)}$ at scale $\mu = m_b$ gives 
\bea
\Delta M_{{\rm B}_s}^{\rm (S)} 
=  \frac{5 f_{B_s}^2 M_{B_s}}{24 M_S^2} ~
{\cal R}e \left[  \frac{7}{5} C_3(\mu) B_3 
- \left( C_4(\mu) B_4 + C_7(\mu) B_7 \right) 
+  \frac{12}{5} \left( C_5(\mu) B_5 + C_8(\mu) B_8 \right) 
\right] \nonumber \\
\label{dMS2}
\eea
with the Wilson coefficients defined as
\bea\label{ScalarCoeff}
& & C_3(\mu) = - 2 r(\mu,M_S )^{-4} ~g_{S 1} g_{S 2} \equiv \overline C_3(\mu)~g_{S 1} g_{S 2}
\nonumber \\
& & C_4(\mu) = - \left[
\left(\frac{1}{2}-\frac{8}{\sqrt{241}} \right) 
r_+ (\mu,M_S) +  
\left(\frac{1}{2}+\frac{8}{\sqrt{241}} \right) 
r_- (\mu,M_S) \right] g_{S 2}^2 \equiv \overline C_4(\mu) ~g_{S 2}^2
\nonumber \\
& & C_5(\mu) = \frac{1}{8 \sqrt{241}} \left[
r_+ (\mu,M_S) - r_- (\mu,M_S)\right] g_{S 2}^2 \equiv \overline C_5(\mu)~ g_{S 2}^2
\label{cs1} \\
& & C_7(\mu)= - \left[
\left(\frac{1}{2}-\frac{8}{\sqrt{241}} \right) 
r_+ (\mu,M_S) +  
\left(\frac{1}{2}+\frac{8}{\sqrt{241}} \right) 
r_- (\mu,M_S) \right] g_{S 1}^2 \equiv \overline C_7(\mu)~g_{S 1}^2
\nonumber \\
& & C_8(\mu) =  \frac{1}{8 \sqrt{241}} \left[
r_+ (\mu,M_S) - r_- (\mu, M_S) \right] g_{S 1}^2 \equiv \overline C_8(\mu)~g_{S 1}^2 \ \ , 
\nonumber 
\eea
where for notational simplicity we have defined 
$r_\pm \equiv r^{(1 \pm \sqrt{241})/6}$.
Note that Eq.~(\ref{dMS2}) is true only for the real spin-0 field $S$. If 
$S$ is a complex field, then only operator $Q_3$ will contribute to Eq.~(\ref{dMS2}).
 
The effective Hamiltonian for the $B^0_s \to \ell^+\ell^-$ decay 
via a heavy scalar S with FCNC interactions is then 
\beq
{\cal H}_{b\to s\ell^+\ell^-}^{\rm (S)} = - \frac{1}{M_S^2} 
\left[
g_{S1}g_{S1}' \widetilde Q_9 +
g_{S1}g_{S2}' \widetilde Q_8 + g_{S1}'g_{S2}  \widetilde Q_3 +
g_{S2}g_{S2}'  \widetilde Q_4 
\right]\ ,
\eeq
and from this, it follows that the branching fraction is 
\bea\label{GammaS}
{\cal B}_{B_s^0 \to \ell^+\ell^-}^{\rm (S)} 
&=& \frac{f_B^2 M_{B_s}^5}
{128\pi m_b^2 M_S^4 \Gamma_{B_s}}
\sqrt{1-\frac{4 m_\ell^2}{M_{B_s}^2}} ~
|g_{S 1} - g_{S 2}|^2 
\nonumber \\
&\times& \
\left[ |g_{S 1}' + g_{S 2}'|^2 
\left(1-\frac{4 m_\ell^2}{M_{B_s}^2}\right) + 
|g_{S 1}' - g_{S 2}'|^2 \right].
\eea
Note that if the spin-0 particle $S$ only has {\it scalar} FCNC couplings, 
{\it i.e.} $g_{S 1} = g_{S 2}$, no contribution to $B_s^0 \to \ell^+\ell^-$
branching ratio is generated at tree level; the non-zero 
contribution to rare decays is instead produced at one-loop level. 
This follows from the {\it pseudoscalar} nature of the $B_s$-meson.

Let us now consider specific models where the correlations 
between the $B_s-\overline{B_s}$ mixing rates and 
(in particular) the 
$B_s \to \mu^+\mu^-$ rare decay can be found. 


\subsection{$Z'$ Boson}
{\it $B_s$ Mixing:} The $B_s$ mixing arising from the $Z'$ pole 
diagram has the same form as in $D^0$ mixing~\cite{Golowich:2007ka}, 
\beq
\Delta M_{B_s}^{\rm (Z')} = {M_{B_s} f_{B_s}^2 
{\hat B}_{B_s} r_1(m_b,M_{Z'}) \over 
3} \cdot {g_{Z's{\bar b}}^2 \over M_{Z'}^2} \ \ , 
\eeq
where $r_1(m_b,M_{Z'})$ is a QCD factor which we take to be 
\beq
r_1(m_b,M_{Z'}) \simeq 0.79 \ \ .
\eeq
This is a compromise between $r_1(m_b,1~{\rm TeV}) = 0.798$ and  
$r_1(m_b,2~{\rm TeV}) = 0.783$.  Solving for the $Z'$ parameters, we
have
\beq
{g_{Z's{\bar b}}^2 \over M_{Z'}^2} = {3 |\Delta M_{B_s}^{\rm (NP)}| \over 
M_{B_s} f_{B_s}^2 {\hat B}_{B_s} r_1(m_b,M_{Z'}) } \le  
3.02 \times 10^{-11}~{\rm GeV}^{-2} 
\eeq
upon using the constraint from $B_s$ mixing.  

{\it $B_s \to \mu^+\mu^-$ Decay:} This has already been calculated for 
$D^0 \to \mu^+\mu^-$ decay in Ref.~\cite{Golowich:2009ii}.  
Inserting obvious 
modifications for $D^0 \to B_s$, we have from 
the branching fraction relation Eq.~(39) of Ref.~\cite{Golowich:2009ii},
\beq\label{GammaZpr}
{\cal B}_{B_s \to \mu^+\mu^-}^{\rm ({Z'})} = 
\frac{G_F f_{B_s}^2 m_\mu^2 M_{B_s}}{16\sqrt{2} \pi \Gamma_{B_s}}
\sqrt{1-\frac{4 m_\mu^2}{M_{B_s}^2}} 
~ {g_{Z's{\bar b}}^2 \over M_{Z'}^2}\cdot {M_Z^2 \over M_{Z'}^2} \ \ .
\eeq
Upon inserting numbers, we obtain 
\beq
{\cal B}_{B_s \to \mu^+\mu^-}^{\rm ({Z'})} \ \le \ 
0.30 \times 10^{-9}~\cdot \left( {1~{\rm TeV} \over M_{Z'}}\right)^2 \ \ .
\eeq
This value is already below the corresponding 
SM prediction (${\cal B}_{B_s \to \mu^+\mu^-}^{\rm (SM)}  = 
3.33 \times 10^{-9}$) even if we take a $Z'$ mass as light as 
$M_{Z'} \simeq 1$~TeV.

\subsection{R Parity Violating Supersymmetry}

One of the models of New Physics that has a rich flavor phenomenology 
is R-parity violating (RPV) SUSY. The crucial difference between studies of 
RPV SUSY contributions to phenomenology of the up-quark 
(see \cite{Golowich:2009ii}) 
and down-type quark sectors is the possibility of tree-level diagrams contributing to 
$B_s$-mixing\footnote{We assume that there is no strong hierarchy between the RPV SUSY couplings 
that favors possible box diagrams.} and $B_s \to \ell^+\ell^-$ 
decays~\cite{Kundu:2004cv,Kao:2009fg,Dreiner:2006gu,Saha:2002kt}.
If one allows for R-parity violation, the following terms should be added to 
the superpotential,
\beq\label{RPVSup}
{\cal W} _{\not R}= \frac{1}{2} \lambda_{ijk} L_i L_j E^c_k +
 \lambda_{ijk}^\prime L_i Q_j D^c_k + 
 \frac{1}{2} \lambda_{ijk}^{\prime\prime} U_i^c D^c_j D^c_k.
\eeq
Here $Q$ and $L$ denote $SU(2)_L$ doublet quark and lepton superfields, and
$U$, $D$ and $E$ stand for the $SU(2)_L$ singlet up-quark, down-quark and charged
lepton superfields. Also, $\{i,j,k\}=1,2,3$ are generation indices. We shall require 
baryon number symmetry by setting $\lambda^{\prime\prime}$ to zero. 
Also, we will assume CP-conservation, so all couplings $\lambda_{ijk}$ and 
$\lambda_{ijk}^\prime$ are treated as real.

\noindent {\it $\BsBsbar$ Mixing}: 
Neglecting the baryon-number violating contribution, the 
Lagrangian describing RPV SUSY contribution to  $\BsBsbar$ mixing can be 
written as
\begin{equation}
{\cal L}_{\not R}=-\lambda_{i23}^\prime \widetilde\nu_{i_L} \overline b_R s_L - 
\lambda_{i32}^\prime \widetilde\nu_{i_L} \overline s_R b_L + h.c.\ \ , 
\end{equation}
where $i=1,2,3$ is a generational index for the sneutrino.  
Matching to Eq.~(\ref{SFCNC}) implies that the
only non-zero contribution comes from the operator $Q_3$. Taking into account 
renormalization group running, we obtain for $\Delta M_s$ 
from the R-parity violating terms,  
\beqa
& & \Delta M_{\rm B_s}^{\rm ({\not R})} = 
\frac{5}{24} f_{B_s}^2 M_{B_s} F(C_{3},B_{3}) 
\sum_i \frac{\lambda_{i23}^\prime \lambda_{i32}^{\prime *}}{M_{\tilde\nu_i}^2},
\eeqa 
where $M_{\tilde\nu_i}$ denotes the mass of the sneutrino of $i$th generation and the function
\beq
F(C_{3},B_{3}) =  \frac{7}{5} \overline C_3(\mu, M_{\tilde\nu_i}) B_3,
\eeq
is defined in terms of reduced Wilson coefficient 
of Eq.~(\ref{ScalarCoeff}) and the B-factor is 
defined in Table~\ref{tab:BagConstants} of the Appendix.

\vspace{0.2cm}
\noindent {\it $B_s \to \mu^+\mu^-$ Decay}: 
In RPV-SUSY, the underlying transition for $B_s \to \mu^+\mu^-$ 
is $s + {\bar b} \to \mu^+ + \mu^-$ via tree-level $u$-squark or sneutrino exchange.
In order to relate the rare decay to the 
mass difference contribution from RPV SUSY 
$\Delta M_{\rm B_s}^{\rm ({\not R})}$, we need to assume that the up-squark contribution is 
negligible. This can be achieved in models where sneutrinos are much lighter than the 
up-type squarks, which are phenomenologically viable. Employing this assumption 
leads to the predicted branching fraction 
\beqa\label{y2} 
{\cal B}^{\rm ({\not R})}_{B_s\to\mu^+\mu^-} &=& 
\frac{f_{B_s}^2 M_{B_s}^3}{64 \ \pi \ \Gamma_{B_s}}
\left(\frac{M_{B_s}}{m_b}\right)^2 \left(1-\frac{2m_\mu^2}{M_{B_s}^2}\right)
\,\sqrt{1-\frac{4m_\mu^2}{M_{B_s}^2}}
\nonumber \\
&\times & 
\left(\left|\sum_i\frac{\lambda_{i22}^*\lambda_{i32}^\prime}{M_{\tilde\nu_i}^2}\right|^2+
\left|\sum_i\frac{\lambda_{i22}\lambda_{i23}^{\prime *}}{M_{\tilde\nu_i}^2}\right|^2 \right). 
\eeqa
In order to relate $B_s \to \mu^+\mu^-$  to $\Delta M_s$ in the framework of 
RPV SUSY, we need to make additional assumptions. In particular, 
we shall assume that the sum is dominated by a single sneutrino state, which we shall
denote by $\tilde\nu_k$. In addition, we will assume that $\lambda_{k23}^\prime=\lambda_{k32}^\prime$,
which will reduce the number of unknown parameters. This assumption is 
not needed,
however, if one wishes to set a bound on a combination of coupling constants directly
from the experimental bound on ${\cal B}_{B_s\to\mu^+\mu^-}$.
Then, neglecting CP-violation,
\beqa\label{y2a} 
{\cal B}^{\rm ({\not R})}_{B_s\to\mu^+\mu^-} = 
k \frac{f_{B_s}^2 M_{B_s}^3}{64 \pi \ \Gamma_{B_s}} 
\left(\frac{\lambda_{i22}\lambda_{i32}^\prime}{M_{\tilde\nu_i}^2}\right)^2
\left(\frac{M_{B_s}}{m_b}\right)^2 \left(1-\frac{2m_\mu^2}{M_{B_s}^2}\right)
\,\sqrt{1-\frac{4m_\mu^2}{M_{B_s}^2}},
\eeqa
where $k=2$ if an assumption that $\lambda_{k23}^\prime=\lambda_{k32}^\prime$ is made, and
$k=1$ otherwise.

\begin{figure} [tb]
\centerline{
\includegraphics[width=9cm,angle=0]{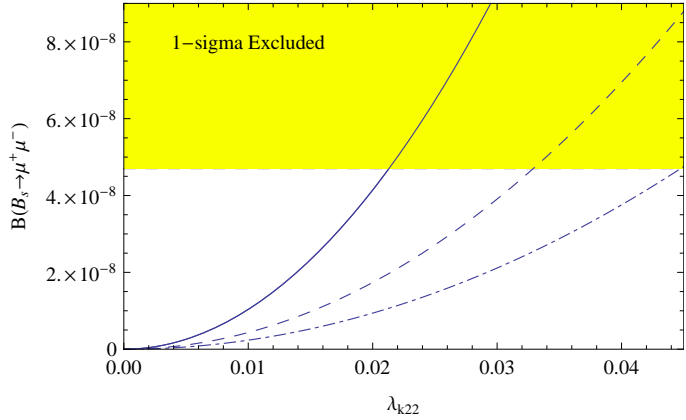}}
\caption{Branching ratio of 
${\cal B}_{B_s^0 \to \mu^+\mu^-}$ as a function
of RPV leptonic coupling $\lambda_{k22}$ and sneutrino mass $M_{\tilde\nu_i} = 100$~GeV,
$150$~GeV, and $200$~GeV (solid, dashed, and dash-dotted lines).
The yellow shaded area represents excluded parameter space.}
\label{RPVGraph}
\end{figure}

Since no $B_s \to \mu^+\mu^-$ signal has yet been seen, we can 
use the experimental bound to obtain an updated constraint on the 
RPV couplings, 
\beq\label{y4}
\lambda_{k22} \lambda^\prime_{k32} \le 
5.5 \times 10^{-6}~\left( \frac{M_{\tilde{\nu}_k}}{100~{\rm GeV}}\right)^2 \ \ .
\eeq
Now, assuming $\lambda_{k23}^\prime=\lambda_{k32}^\prime$, one can relate the branching ratio 
${\cal B}_{B_s\to\mu^+\mu^-}$ to $x_{B_s}^{(\not R)}$,
\beq\label{relate3}
{\cal B}^{\rm ({\not R})}_{B_s \to \mu^+\mu^-} =
\frac{3}{20 \pi} 
\frac{M_{B_s}^2}{F(C_{3},B_{3})}
\left(\frac{M_{B_s}}{m_b}\right)^2 
\left(1-\frac{2m_\mu^2}{M_{B_s}^2}\right)
\sqrt{1-\frac{4m_\mu^2}{M_{B_s}^2}} \ x_{B_s}^{(\not R)}  \
\frac{\lambda_{k22}^2}{M_{\tilde\nu_i}^2}.
\eeq
It is possible to plot the dependence of ${\cal B}_{B_s\to\mu^+\mu^-}$ on 
$\lambda_{k22}$ for different values of $M_{\tilde\nu_i}$, 
which we present in Fig.~\ref{RPVGraph}.
 
\subsection{Family (Horizontal) Symmetries}

The gauge sector in the Standard Model has a large global symmetry which
is broken by the Higgs interaction~\cite{Sher:1988mj}.  
By enlarging the Higgs sector,
some subgroup of this symmetry can be imposed on the full SM lagrangian
and the symmetry can be broken spontaneously.  This family symmetry can be
global~\cite{fcnch} as well as gauged~\cite{Maehara:1977nq}.
If the new gauge couplings are very weak
or the gauge boson masses are large, the difference between a gauged or
global symmetry is rather difficult to distinguish 
in practice~\cite{begmemor}.  In 
general there would be FCNC effects from both the gauge and scalar
sectors.  Here we study the gauge contribution.  Consider the family
gauge symmetry group $SU(3)_G$ acting on the three left-handed
families.  Spontaneous symmetry breaking renders all the gauge bosons
massive.  If the SU(3) is broken first to SU(2) before being
completely broken, we may have an effective `low' energy 
symmetry $SU(2)_G$.  This means that the gauge bosons 
${\bf G} \equiv \left\{ G_i \right\} \ 
(i = 1,\dots ,3)$ are much lighter than the 
$\left\{ G_k \right\} \ (k = 4,\dots ,8)$.  For simplicity 
we assume that after symmetry breaking the gauge boson mass matrix 
is diagonal to a good approximation.  If so, the light gauge bosons 
${\bf G}$ are mass eigenstates with negligible mixing.

The LH doublets 
\begin{equation}
\left( \begin{array}{c} u^0\\ d^0 \end{array} \right)_L \ , \qquad 
\left( \begin{array}{c}  c^0 \\  s^0
\end{array} \right)_L  \ , \qquad 
\left( \begin{array}{c}  t^0 \\  b^0
\end{array} \right)_L  \quad   \quad ,
\end{equation}
transform as $I_G = 1/2$ under $SU(2)_G$, as do the lepton doublets
\begin{equation}
\left( \begin{array}{c} \nu^0_e \\ e^0 \end{array} \right)_L \ ,\qquad 
\left( \begin{array}{c} \nu^0_\mu \\ \mu^0 \end{array} \right)_L \qquad 
\left( \begin{array}{c}  \nu^0_\tau \\  \tau^0
\end{array} \right)_L \ \ .  
\end{equation}
and the right-handed fermions are singlets under $SU(2)_G$.   
In the above, the superscript $`o'$ 
refers to the fact that these are weak eigenstates and not
mass eigenstates.  The couplings of fermions to the light 
family gauge bosons ${\bf G}$ is given by
\begin{equation}
L = f \left[ \bar{\psi}_{d^{0},L} \gamma_{\mu}
\bm{\tau}  \cdot {\bf G}^\mu \psi_{{d{^0},L}} + 
\bar{\psi}_{u^{0},L} \gamma_{\mu}
  {\bm\tau} \cdot {\bf G}^\mu \psi_{{u{^0},L}} + 
\bar{\psi}_{\ell^{0},L} \gamma_{\mu}
  {\bm\tau} \cdot {\bf G}^\mu \psi_{{\ell{^0},L}} \right] \ \ , 
\end{equation}
where $f$  denotes the coupling strength and ${\bm\tau}$ are the
generators of $SU(2)_G$

The fermion mass eigenstates are given by, first for quarks, 
\begin{equation}
\left( \begin{array}{c} d \\ s \\ b \end{array} \right)_L 
 \ = \ U_d 
\left( 
\begin{array}{c}  
d^0 \\  s^0 \\ b^0
\end{array} \right)_L 
\qquad {\text and} \qquad 
\left (
\begin{array}{c}
u \\ c \\  t 
\end{array}  \right)_L \ = \ U_u  
\left (
\begin{array}{c}
u^0 \\  c^0  \\  t^0
\end{array} \right )_L
\label{qks}
\end{equation}
and then for leptons, 
\begin{equation}
\left (
\begin{array}{c}
e \\ \mu \\ \tau 
\end{array} \right )_L \ = \ U_\ell
\left (
\begin{array}{c}
u^0 \\ \mu^0 \\ \tau^0
\end{array} \right )_L
\qquad \text{and} \qquad 
\left (
\begin{array}{c}
\nu_1 \\ \nu_2 \\ \nu_3 
\end{array} \right )_L \ = \  U_\nu
\left (
\begin{array}{c}
\nu_e^0 \\ \nu_\mu^0 \\ \nu_\tau^0
\end{array} \right )_L \ \ .
\label{u}
\end{equation}
The four matrices $U_d, U_u, U_\ell$ and $U_\nu$ are unknown, except for 
\begin{equation}
U_u^\dagger U_d = V_{\rm CKM} \qquad \text{and} \qquad  
U_\nu^\dagger U_\ell \ = V_{\rm MNSP} \ \ .
\end{equation}
where $V_{\rm MNSP}$ is the Maki-Nakagawa-Sakata-Pontcorvo lepton 
mixing matrix.  
The couplings of the gauge bosons relevant for the $B_s$ system in the
mass basis are:
\begin{eqnarray}
& & L=  f \bigg[  G_1^\mu . \left( U_{b1} U^*_{s2} \bar{b}_L
  \gamma_\mu s_L  \ + \  U_{s1} U^*_{b2}  \bar{s}_L 
\gamma_\mu b_L + U_{b2}U^*_{s1} \bar{b}_L \gamma_\mu s_L  
+ U_{s2}  U^*_{b1}   \bar{s}_L\gamma_\mu b_L \right)  \nonumber \\
& & + i G_2^\mu\left( - U_{b1} U^*_{s2} \bar{b}_L \gamma_\mu s_L \ - \
  U_{s1} U^*_{b2} \bar{s}_L \gamma_\mu b_L  
+  \ U_{b2} U^*_{s1} \bar{b}_L \gamma_\mu s_L \ + \ 
U_{s2} \ U^*_{b1} \bar{s}_L \gamma_\mu b_L \right)   \nonumber \\
& & +  G_3^\mu \left(  U_{b1} U^*_{s1} \bar{b}_L \gamma_\mu s_L \ + \ 
U_{s1}   U^*_{b1} \bar{s}_L \gamma_\mu b_L 
- \ U_{b2} U^*_{s2}  \bar{b}_L \gamma_\mu \bar{s}_L \ 
- \ U_{s2} U^*_{b2} \bar{s}_L \ \gamma_\mu \ b_L \right)  \bigg]
\end{eqnarray}
The contribution to $B_s^0 - \bar{B}_s^0$ mixing is given by
\begin{equation}
\Delta M_{B_{s}}^{\rm (FS)}  = \frac{2 M_{B_s} f^2_{B_s} B_{B_s} 
r (m_{{B_s}, M})} {3 } \ f^2 \left[
\frac{A}{m_1^2} \ + \frac{C}{m_3^2}  +  \frac{B}{m_2^2} \right]
\label{famsymm}
\end{equation}
where 
\begin{eqnarray}
\begin{array}{l}
A =  {\rm Re}\left[ \left ( U_{b1} \ U^*_{s2} + U_{b2} 
\ U^*_{s1} \right )^2\right]
                \\
B = -{\rm Re}\left[\left ( U_{b1} \ U^*_{s2} - \ U_{b2} \ U^*_{s1} 
\right )^2\right]
\label{abc} \\
C = {\rm Re}\left[\left ( U_{b1} U^*_{s1} - U_{b2}U^*_{s2}\right)^2\right] 
\end{array}
\end{eqnarray}

In a simple scheme of symmetry breaking~\cite{Monich:1980rr}, 
one obtains $m_1 = m_3$ and the square bracket in Eq.~(\ref{famsymm}) 
becomes
\begin{equation}
\left [\frac{A+C}{m^2_1} \ \ + \frac{B}{m^2_2}  \right ] \ \ .
\end{equation}
Although the matrices $U_i\ (i = d,u,\ell)$ in principle are unknown,
it has been argued that a reasonable ansatz~\cite{q7}, 
which is incorporated in many models is 
$U_u  = I, ~ U_d^\dagger = V_{\rm CKM}$.
In this case\footnote{Here, we use values listed in Ref.~\cite{PDG}.} 
one can simplify $A,B$ and $C$ further:
\begin{equation}
A,B \ll C \simeq 1.6 \times 10^{-3} \ \ .
\end{equation}
Thus the $B_s$ mixing becomes 
\beq
\Delta M_{B_s}^{\rm (FS)} \simeq \frac{ 2 M_{{B_S}} f^2_{B_{s}} 
{\hat B}_{B_s} r(m_b,M) }{ 3 }~ {f^2 \over m_1^2} 
~1.6\times 10^{-3} \ \ , 
\label{fammix}
\eeq
so that, substituting experimental bound 
$\Delta M_{B_s}^{\rm (FS)}=\Delta M_{B_s}^{\rm (NP)}$,
\beq
{f^2 \over m_1^2} \leq \frac{3 |\Delta M_{B_s}^{\rm (NP)}|}
{2 M_{{B_S}} f^2_{B_{s}} {\hat B}_{B_s}
r(m_b,M) 1.6\times 10^{-3}}\ \ .
\label{fm1}
\eeq
The same above ansatz also implies that $U_\ell^\dagger = 
U_{\rm MNSP}$ and $U_\nu=1$.  
Then the coupling of the gauge bosons to muon pairs is given by
\begin{eqnarray}
& & {\cal L}_{{\rm G}\mu^+\mu^-} =  f \bigg[ \left (U^*_{\mu 1} 
U_{\mu 2}
 + U_{\mu 1}\ U^*_{\mu 2}  \right )G_1^\lambda 
\nonumber \\
& &  +  \ i  \left(  - U_{\mu 1} \ U^*_{\mu 2}
 + U^*_{\mu 1}U_{\mu 2} \right ) G_2^\lambda
\ + \left(U_{\mu 1} \ U^*_{\mu 1}
 - U_{\mu 2}\ U^*_{\mu 2} \right ) G_3^\lambda
\bigg] \bar{\mu}_L \gamma_\lambda \mu_L \ \ .
\label{gmm}
\end{eqnarray}
The branching ratio for $B_s \rightarrow \mu^+\mu^-$ is given by
\begin{eqnarray}
& & {\cal B}_{B_s \to \mu^+ \mu^-} = \frac{ M_{{B_S}} f^2_{B_{s}} \
  m_\mu^2 }{ 32 \pi \Gamma_{B{_s}} }  \ f^4  
\bigg| \frac { \left ( U_{b1}U^*_{s2} \ + U_{b2}U^*_{s1}\right )  
\left ( U_{\mu 1} U^*_{\mu 2} + U^*_{\mu 1} 
\ U_{\mu 2}  \right )} {m_1^2} \nonumber \\
& & - \frac{ \left ( U_{b1} U^*_{s2} - U_{b2}U^*_{s1}
  \right )  \left ( U_{\mu 1} U^*_{\mu 2} - U_{\mu 2} 
U^*_{\mu 1} \right )}{m^2_2} 
+ \frac{ \left ( U_{b1} U^*_{s1} - U_{b2} U^*_{s2}
  \right )  \left ( U_{\mu 1} U^*_{\mu 1} - U_{\mu 2} 
U^*_{\mu 2} \right )} {m^2_3} \bigg|^2
\end{eqnarray}
Next we employ the approximation (well-supported empirically) 
that $U_{\rm MNSP} \simeq U_{\rm TBM}$, where 
$U_{\rm TBM}$ is the tri-bi-maximal matrix~\cite{Harrison:2002er}.
Then Eq.~(\ref{gmm}) becomes 
\begin{eqnarray}
 {\cal L}_{{\rm G}\mu^+\mu^-} = - f 
\left [ \frac{\sqrt{2}}{3} ~G_1^\mu + \frac{1}{6} ~
G_3^\mu   \right ] \bar{\mu}_L \gamma_\mu \mu_L \ \ .
\end{eqnarray}
With this, the contribution to the branching ratio for $B_s \to
\mu^+ \mu^-$ becomes
\begin{eqnarray}
B_{B_s \to \mu^+\mu^-} &=& \frac{M_{B{_s}} f^2_{B_s} 
m_\mu^2 f^4}{32 \pi \Gamma_{B_{s}}}
\left [ \frac{\sqrt{2}}{3} \left(1.1\times 10^{-2}\right) \ + \ 
\frac{1}{6} \times 0.04 \right ]^2 {1 \over m_1^4} \nonumber \\
&\simeq& \frac{M_{B_s} f^2_{B_s} 
    m_\mu^2 f^4}{32 \pi \Gamma_{B_{s}}}~
{1.4 \times 10^{-4} \over m_1^4}  \ \ .
\label{brmumu}
\end{eqnarray}
The dependence on unknown factors in Eq.~(\ref{brmumu}) 
({\it i.e.} $(f/m_1)^4$) 
can be entirely removed by using the bound in 
Eq.~(\ref{fm1}) to yield 
\beqa
& & {\cal B}_{B_s \to \mu^+ \mu^-}^{\rm (FS)} \le \frac{3.85 m_\mu^2 }
{\pi M_{B_S}  \Gamma_{B_s} 
\left( f_{B_s} {\hat B}_{B_s} r(m_b,m_1) \right)^2 }
~|\Delta M_{B_s}^{\rm (NP)}|^2 \ \ .
\label{fsbr}
\eeqa
From the bounds of Eqs.~(\ref{av-np}),(\ref{av-onesig}), we obtain 
%
\beqa
& & {\cal B}_{B_s \to \mu^+ \mu^-}^{\rm (FS)} \le 
0.92 \times 10^{-12} \ \ .
\eeqa


\subsection{FCNC Higgs interactions}

Many extensions of the Standard Model contain multiple scalar 
doublets, which increases the
possibility of FCNC mediated by flavor non-diagonal interactions of
neutral components. While
many ideas exist on how to suppress those interactions 
(see, {\it e.g.}~\cite{Hall:1993ca,Cheng:1987rs,Pich:2009sp}), 
the ultimate test of those ideas would involve direct 
observation of scalar-mediated FCNC.

Consider a generic Yukawa interaction consisting of a 
set of $N$ Higgs doublets $H_n$ $(n=2, .., N)$ with SM fermions,
\beq
{\cal H}_Y = \lambda_{ijn}^U \overline Q_{Li} U_{Rj} \widetilde H_n 
+  \lambda_{ijn}^D \overline Q_{Li} D_{Rj} H_n 
+  \lambda_{ijn}^E \overline L_{Li} E_{Rj} H_n + \mbox{h.c.}\ \ ,
\label{int1}
\eeq
where $\widetilde H_n =  i \sigma_2 H_n^*$ and $Q_{Li}$ 
($L_{Li}$) are respectively the left-handed weak 
doublets of an $i$th-generation of quarks (leptons). 
Restricting the discussion to $B_{s}$ Mixing and $B_{s} \to 
\mu^+\mu^-$ decay, we find that Eq.~(\ref{int1}) reduces to 
\beq\label{FCNCHiggs}
{\cal H}^{H}_Y = \lambda_{23n}^D  \overline s_L b_R  \Phi^0_n \ + 
 \lambda_{32n}^D  \overline b_L s_R  \Phi^0_n +
\lambda_{22n}^E  \overline \mu_L \mu_R \Phi^0_n + \mbox{h.c.},
\eeq
where $\Phi^0_n \equiv \left(\phi_n^0+i a_n^0\right)/ \sqrt{2}$. 
Bringing this to the form of Eq.~(\ref{SFCNC}) 
and confining the discussion only to the contribution 
of the lightest $\phi_n^0$ and $a_n^0$ states, we obtain 
\bea\label{FCNCHiggs2}
{\cal H}^{H}_Y &=& \frac{\lambda_{23}^{D\dagger}}{\sqrt{2}} 
\ \overline b_R s_L  \phi^{0} \ + 
 \frac{\lambda_{32}^D}{\sqrt{2}}  \ \overline b_L s_R  \phi^0 +
\frac{\lambda_{22}^E}{\sqrt{2}}  \ \overline \mu_L \mu_R \phi^0
\nonumber \\
&-&  i \frac{\lambda_{23}^{D\dagger}}{\sqrt{2}} \ \overline b_R s_L  a^{0} \ + 
i \frac{\lambda_{32}^D}{\sqrt{2}} \ \overline b_L s_R  a^0 + \
i \frac{\lambda_{22}^E}{\sqrt{2}} \ \overline \mu_L \mu_R a^0 
+ ... \ + \mbox{h.c.}\ \ ,
\eea
where ellipses stand for the terms containing heavier $\phi_n^0$ and 
$a_n^0$ states whose contributions
to $\Delta M_{B_s}$ and ${\cal B}_{B_s \to \mu^+\mu^-}$ will be suppressed.

If the matrix of coupling constants in Eq.~(\ref{FCNCHiggs2}) 
is Hermitian, {\it e.g.} 
$\lambda_{23}^{D\dagger} = \lambda_{32}^D$, then we can 
identify the couplings of  Eq.~(\ref{SFCNC}) as
\bea\label{SCouplings}
& & g_{S_1} = g_{S_2} =   \frac{\lambda_{32}^D}{\sqrt{2}}, \quad\quad
g_{S_1}^\prime  = g_{S_2}^\prime =   \frac{\lambda_{22}^E}
{\sqrt{2}}\ \  
\eea
for {\it scalar}  interactions and 
\bea\label{PCouplings}
& & g_{S_1} = - g_{S_2} =   \frac{i \lambda_{32}^D}{\sqrt{2}}, \quad
g_{S_1}^\prime  = - g_{S_2}^\prime =  \frac{i \lambda_{22}^E}{\sqrt{2}}
\ \ 
\eea
for {\it pseudoscalar} interactions.

To proceed, we need to separate two cases: (i) the 
lightest FCNC Higgs particle is a scalar, and
(ii) the lightest FCNC Higgs particle is pseudoscalar.

\subsubsection{Light scalar FCNC Higgs}

The case of relatively light scalar Higgs state is quite common, 
arising most often in Type-III two-Higgs doublet models (models without
natural flavor conservation)~\cite{Barger:1989fj,Atwood:1996vj,Blechman:2010cs}. 

\vspace{0.2cm} 
\noindent{\it $B_s^0$-${\bar B}_s^0$ Mixing}: 
Given the general formulas of Eq.~(\ref{dMS2}), it is easy to
compute the contribution to $\Delta M_{{\rm B}_s}^{\rm (\phi)}$ 
of an intermediate scalar ($\phi$) with FCNC couplings,  
\bea
\Delta M_{{\rm B}_s}^{\rm (\phi)} 
&=&  \frac{5 f_{B_s}^2 M_{B_s}f_\phi(\overline C_i,m_b)}
{48 } ~ \left( {\lambda_{32}^D \over M_\phi} \right)^2 
\ \ , \label{dMH2} \\
f_\phi(\overline C_i,m_b) &\equiv& 
\frac{7}{5} \overline C_3(m_b) B_3 
- \left( \overline C_4(m_b) B_4 + \overline C_7(m_b) B_7 \right) 
+  \frac{12}{5} \left(\overline C_5(m_b) B_5 + 
\overline C_8(m_b) B_8 \right) \ , \nonumber 
\eea
with 'reduced' Wilson coefficients 
$\{ \overline C_i(\mu)\}$ given in Eq.~(\ref{cs1}).

\vspace{0.2cm} 
\noindent {\it $B_s^0 \to \mu^+\mu^-$ Decay}: 
Comparing Eq.~(\ref{SCouplings}) to Eq.~(\ref{GammaS}), we can 
easily see that the branching fraction for the 
rare decay $B_s^0\to \ell^+\ell^-$ is zero for the intermediate scalar Higgs,
\beq
{\cal B}_{B_s^0\to \ell^+\ell^-}^{\rm (\phi)} = 0\ \ .
\eeq
This is consistent with what was already discussed in 
Sec.~\ref{sec:TreeLevel} and
implies that the FCNC Higgs model does not produce a contribution to 
$B_s^0\to \mu^+\mu^-$ at tree level. The non-zero contribution to
$B_s^0\to \mu^+\mu^-$ decay is produced at one-loop level~\cite{Diaz:2004mk}.

\subsubsection{Light pseudoscalar FCNC Higgs}

The case of a lightest {\it pseudoscalar} Higgs state 
can occur in the non-minimal supersymmetric
standard model (NMSSM)~\cite{Nilles:1982dy,Ellis:1988er,
Ellwanger:1996gw,Hiller:2004ii}
or related models~\cite{Dobrescu:1999gv}. In NMSSM, 
a complex singlet Higgs is
introduced to dynamically solve the $\mu$ problem. The resulting pseudoscalar
can be as light as tens of GeV. 
This does not mean, however, that it necessarily gives the 
dominant contribution to both $B_s^0-\overline B_s^0$
mixing and the $B_s^0 \to \mu^+\mu^-$ decay rate since there can be loop 
contributions from other Higgs states. In the following, 
we shall work in the region of the parameter space where it does.

\vspace{0.2cm} 
\noindent{\it $B_s^0$-${\bar B}_s^0$ Mixing}: 
The contribution to $\Delta M_{{\rm B}_s}^{\rm (a)}$ due to
intermediate pseudoscalar with flavor-changing
couplings can be computed using the general formula
in Eq.~(\ref{dMS2}) along with the identification given in 
Eq.~(\ref{PCouplings}), 
\bea\label{dMa}
\Delta M_{{\rm B}_s}^{\rm (a)} 
&=&  \frac{5 f_{B_s}^2 M_{B_s} f_a(\overline C_i,m_b)}{48} 
\left({\lambda_{32}^D \over M_a} \right)^2\ \ ,
\\
f_a(\overline C_i,m_b) 
&=& \left[  \frac{7}{5} \overline C_3(m_b) B_3 
+ \left( \overline C_4(m_b) B_4 + \overline C_7(m_b) B_7 \right) 
-  \frac{12}{5} \left(\overline C_5(m_b) B_5 + 
\overline C_8(m_b) B_8 \right) 
\right]  \nonumber 
\eea
with `reduced' Wilson coefficients $\overline C_i(\mu)$ again 
being defined in Eq.~(\ref{cs1}).

\begin{figure} [tb]
\centerline{
\includegraphics[width=9cm,angle=0]{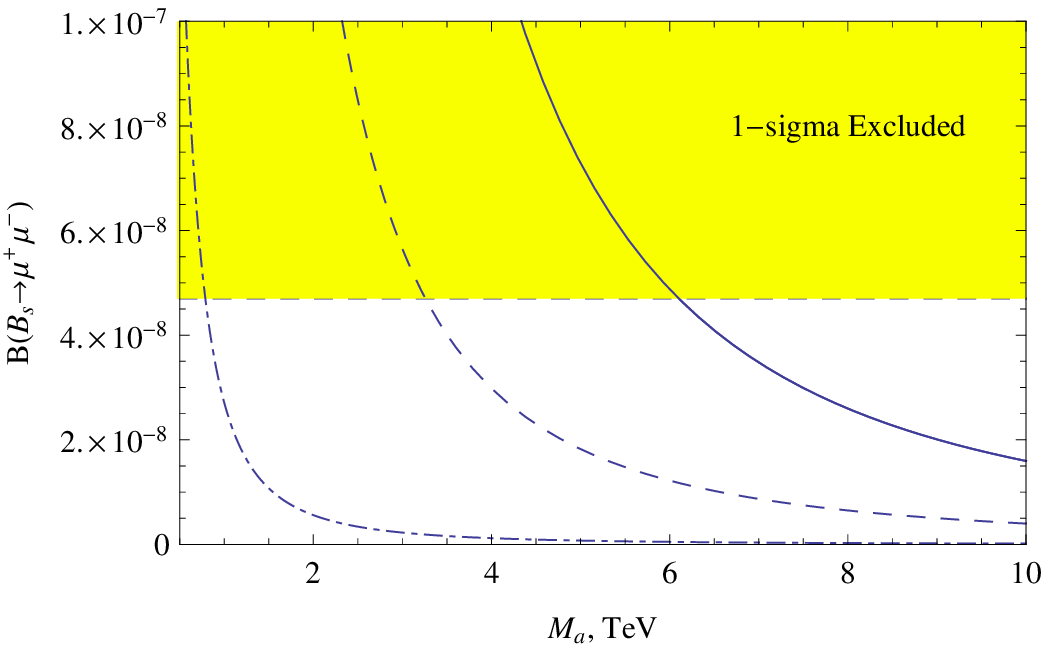}
\includegraphics[width=9cm,angle=0]{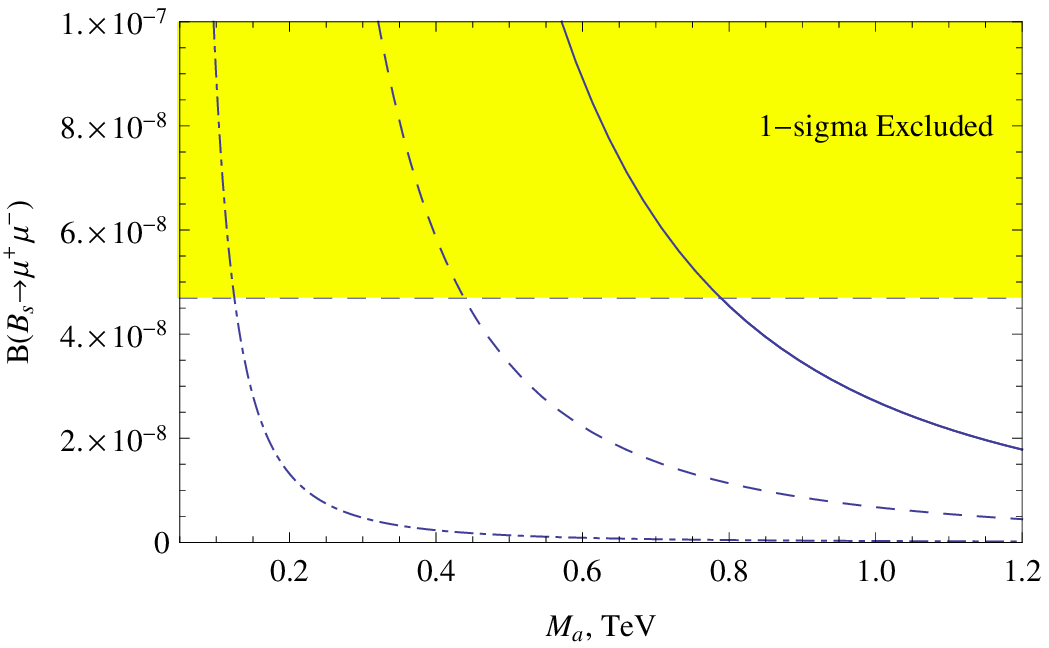}}
\caption{Branching ratio of 
${\cal B}_{B_s^0 \to \mu^+\mu^-}$ as a function
of pseudoscalar Higgs mass $M_a$. Left: $\lambda_{22}^E=1,0.5,0.1$ 
(solid, dashed, dash-dotted lines).
Right: $\lambda_{22}^E=0.1,0.05,0.01$ 
(solid, dashed, dash-dotted lines).
In each figure, the yellow shaded area represents excluded parameter 
space.}
\label{aGraph}
\end{figure}

\vspace{0.2cm} 
\noindent {\it $B_s^0 \to \mu^+\mu^-$ Decay}: 
The branching ratio for rare decay can be computed with the help of
the general formula of Eq.~(\ref{GammaS}),
\bea\label{GammaA}
{\cal B}_{B_s^0 \to \ell^+\ell^-}^{\rm (a)} 
&=& \frac{1}{32 \pi} \frac{f_B^2 M_{B_s}^5 }
{m_b^2  \Gamma_{B_s}}
\left(1-\frac{4 m_\ell^2}{M_{B_s}^2}\right)^{1/2}
\left({\lambda_{32}^D ~\lambda_{22}^E \over M_a^2} \right)^2 \ \ .
\eea
We can now eliminate one of the three unknown 
parameters ($\lambda_{32}^D$,
$\lambda_{22}^E$, and $M_a$) which appear 
in Eqs.(\ref{dMa}) and (\ref{GammaA}). 
We choose to eliminate $\lambda_{32}^D$, so
\bea\label{aRel}
{\cal B}_{B_s^0 \to \ell^+\ell^-}^{\rm (a)} 
&=& \frac{3}{10 \pi} \cdot 
\frac{M_{B_s}^4 x_{s}^{\rm (a)}}{m_b^2 f_a(\overline C_i,m_b)}
\left(1-\frac{4 m_\ell^2}{M_{B_s}^2}\right)^{1/2}
\left(\frac{\lambda_{22}^E}{M_a}\right)^2  \ \ ,
\eea
where $x_{s}^{\rm (a)} =\Delta M_{{\rm B}_s}^{\rm (a)} /\Gamma_{B_s}$. 
As one can see, the unknown factors enter Eq.~(\ref{aRel}) 
in the combination $\lambda_{22}^E/M_a$.
It is, however, more convenient to plot the dependence on 
$M_a$ for different values of $\lambda_{22}^E$, 
which we present in Fig.~\ref{aGraph}.
 
It must be emphasized that the discussion above assumed the absence of large destructive interference
of the NP and SM contributions to $\BsBsbar$ mixing. Concrete models where such interference is present 
(and thus the New Physics contribution is larger than the SM one) can be constructed~\cite{Endo:2010yt}. 
In such models possible contribution to $B_s \to \mu^+\mu^-$ could be large.

\subsection{Fourth generation models}

One of the simplest extensions of the Standard Model involves addition of the 
sequential fourth generation of chiral quarks~\cite{Holdom:2009rf,Buras:2010pi,Hou:2010mm}, 
denoted for the lack of the better names by $t'$ and $b'$.  The addition of the 
sequential fourth generation of quarks 
leads to a 4$\times$4 CKM quark mixing matrix~\cite{Chanowitz:2009mz}. This implies 
that the parameterization 
of this matrix requires six real parameters and three phases. Besides providing
new sources of CP-violation, the two additional phases can affect the branching ratios 
considered in this paper due to interference effects~\cite{Bobrowski:2009ng}. 

There are many existing constraints on the parameters related to the fourth generation of 
quarks. In particular, a fit of precision electroweak data 
(S and T parameters)~\cite{Novikov:1994zg,Novikov:2001md,Erler:2010sk} implies
that the masses of the new quarks are strongly constrained to be~\cite{Kribs:2007nz}
\beq
m_{t'}-m_{b'} \simeq \left(1+\frac{1}{5} \frac{m_H}{(115\mbox{~GeV})} \right) \times 50\mbox{~GeV},
\eeq
with $m_{t'} > 400$~GeV. Here $m_H$ is the SM Higgs mass, which we take for simplicity to be
120 GeV. We also used updated constraints on CKM matrix elements~\cite{Alok:2010zj}.

\begin{figure} [tb]
\centerline{
\includegraphics[width=9cm,angle=0]{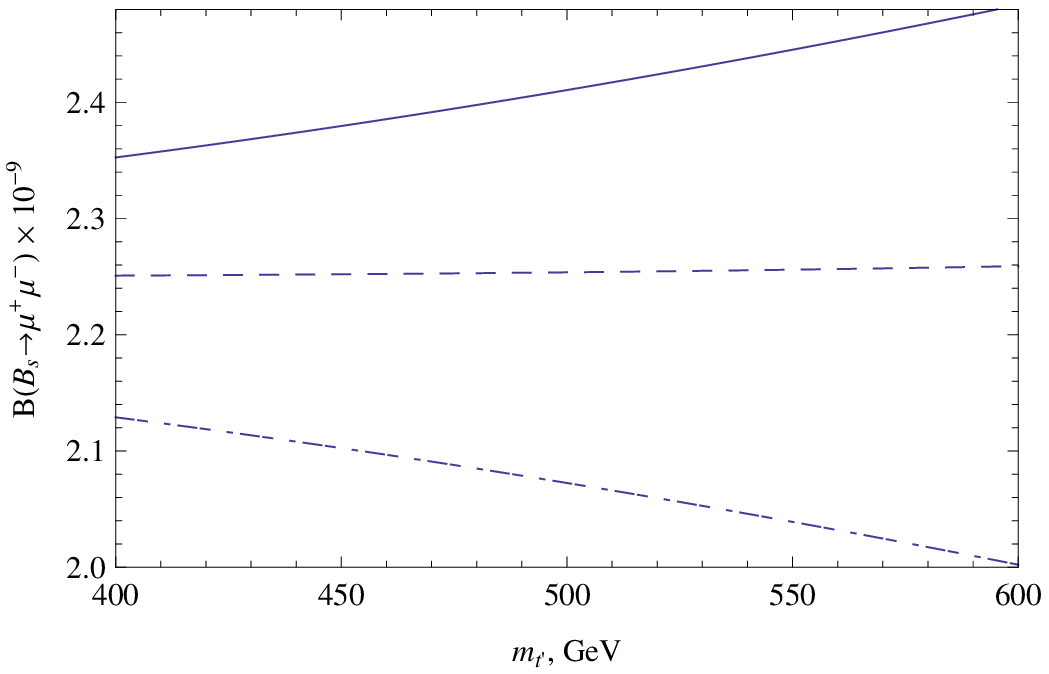}
\includegraphics[width=9cm,angle=0]{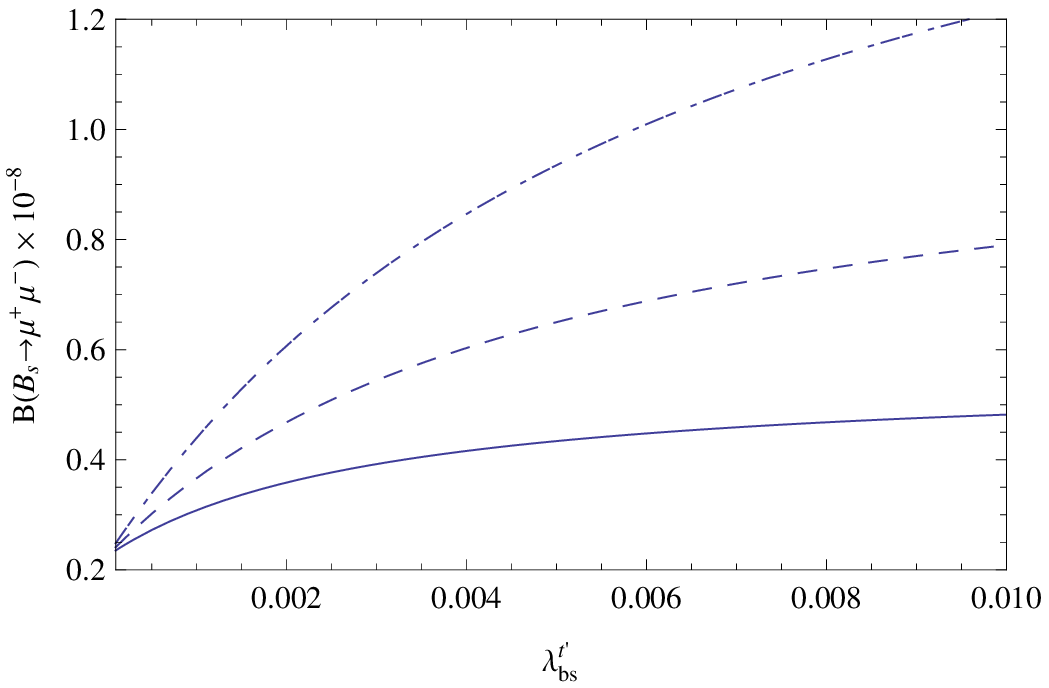}}
\caption{Left: branching ratio of 
${\cal B}_{B_s^0 \to \mu^+\mu^-}$ as a function
of the top-prime mass $m_{t'}$ for different values of the phase $\phi_{t's}=0,\pi/2,\pi$ 
(solid, dashed, dash-dotted lines) and 
$\lambda_{bs}^{t'}=|V_{t's} V_{t'b}^*|\simeq 10^{-4}$~\cite{Alok:2010zj} 
(see also~\cite{Nandi:2010zx}).
Right: branching ratio of ${\cal B}_{B_s^0 \to \mu^+\mu^-}$ as a function
of the CKM parameter combination $\lambda_{bs}^{t'}$ with $\phi_{t's}=0$ and different 
values of $m_{t'}=400$~GeV (solid), $500$~GeV (dashed), and $600$~GeV (dash-dotted).}
\label{FourGenGraph}
\end{figure}

The relationship between $\Delta M_{B_s}$ and ${\cal B}_{B_s \to \mu^+\mu^-}$ in the
model with four generations of quarks has been previously studied in detail in \cite{Soni:2010xh}. 
Here we update their result. The branching ratio of $B_s \to \mu^+\mu^-$
can be related to the experimentally-measured\footnote{Here we 
use $\Delta M_{B_s}$ from Table~\ref{tab:corr}, as the separation of NP and SM contributions used 
in the rest of this paper, $x_{B_s}=x_{SM3}+x_{SM4}$, is not possible
due to loops with both $t'$ and $t$, $c$, or $u$ quarks.}  $x_{B_s}$ as~\cite{Soni:2010xh}
\beq
{\cal B}_{B_s \to \mu^+\mu^-} = \frac{3 \alpha^2 m_\mu^2 x_{B_s}}{8\pi \hat B_{B_s} M_W^2} 
\sqrt{1-\frac{4 m_\mu^2}{m_{B_s}^2}} \frac{\left|C_{10}^{tot}\right|^2}{\left|\Delta^\prime\right|},
\eeq
where the parameter $\Delta^\prime$ is a $B_s$-mixing loop parameter~\cite{Soni:2010xh},
\beq
\Delta^\prime = \eta_t S_0(x_t) + \eta_{t'} R_{t't}^2 S_0(x_{t^\prime}) +
2 \eta_{t^\prime}  R_{t't} S_0(x_t,x_{t^\prime}),
\eeq
and $R_{t't}=V_{t's} V_{t'b}^*/V_{ts} V_{tb}^*$. $\hat B_{B_s}$ can be obtained from Table~\ref{tab:corr}.
The definition of the function $S_0(x_t,x_{t^\prime})$ can be found in Ref.~\cite{Soni:2010xh}. 
The Wilson coefficient $C_{10}^{tot}$ is defined as
\beq
C_{10}^{tot}(\mu) = C_{10} (\mu) + R_{t't} C_{10}^{t'}(\mu)
\eeq
with $C_{10}^{t'}$ obtained by substituting $m_{t'}$ into the SM expression for $C_{10}$~\cite{Buras:1994dj}.
The results can be found in Fig.~\ref{FourGenGraph}. As one can see, the resulting branching ratios are
still lower than the current experimental bound of Eq.~(\ref{bmm}), but for the values of the 
four-generation CKM matrix $\lambda_{bs}^{t'}=|V_{t's} V_{t'b}^*|$ of about $0.01$, disfavored by~\cite{Alok:2010zj}, 
but still favored by~\cite{Nandi:2010zx}, can be quite close to it.

\section{Conclusion}

Experiment has determined $\Delta M_{B_s}$ exceedingly well.  
The Standard Model determination provides a consistent value, although 
with a markedly greater uncertainty (due mainly to the dependence on 
the nonperturbative quantity $f^2_{B_s}{\hat B}_{B_s}$ and to 
a lesser extent on the CKM mixing element $V_{ts}$). We have argued 
that this fact can be used  to constrain NP predictions for other processes, 
such as the $B_{s} \to \mu^+\mu^-$ transition considered here.

We expect this kind of correlation to be a rather general feature of 
New Physics models, provided there is an overlap between the 
NP parameters which describe $\Delta M_{B_s}$ and (for our 
purposes here) $B_{s} \to \mu^+\mu^-$.  
However, given the abundance of New Physics scenarios, each with 
its particular structure, it is not reasonable to expect any 
{\it universal} correlation between $B_s$-mixing and $B_s \to \mu^+\mu^-$.
Instead, what we have done in this paper is to analyze several NP models 
in detail.  In each case, we have first determined the set of 
unknown NP parameters and then, using dynamical assumptions, have 
been able to reduce (or entirely eliminate) the arbitrariness.
Analyzing specific NP models this way has two purposes: 
to serve as an instructive example for further study and 
to see what kinds of numerical predictions these particular models yield.  

Not surprisingly, the simplest model (with a single $Z'$ boson) 
provides a strong correlation between $\Delta M_{B_s}$ and 
$B_{s} \to \mu^+\mu^-$ in which the latter is determined 
in terms of $M_{Z'}$.  An even stronger prediction occurs in 
the particular version of the Family Symmetry model discussed 
earlier, where a clean determination of $B_{s} \to \mu^+\mu^-$ 
is obtained.  In this instance, a set of reasonable assumptions 
allows for the initial presence of unknown parameters to be 
totally overcome.  A similar, but not quite as fortunate, situation 
occurs for 
R-parity violating supersymmetry, wherein a reasonable assumption 
partially reduces the NP parameter set.  
In this case, $B_{s} \to \mu^+\mu^-$ can be expressed in terms 
of a ratio of a coupling constant and sneutrino mass $M_{\tilde{\nu}}$. 
The flavor-changing Higgs model 
turns out to be less accommodating in that no set of assumptions 
known to us can reduce the original set of three unknown parameters.
Thus, the constraint from $B_s$ mixing still leaves one with 
two unknowns (see Fig.~\ref{aGraph}). We also updated constraints on
the models with fourth sequential generation of quarks.
Of course, additional NP models are available for study,  
{\it e.g.} R-parity conserving supersymmetry~\cite{Future}, and work 
proceeds on these.

Finally, as discussed in Sect. III, it would be of interest 
to address the impact of NP CP-violating contributions to 
$B_s$ mixing.  Indeed, we plan do so in a future project, 
but first await more accurate data on $\Delta \Gamma_s$ or
studies of $B_s \to J/\psi \phi$ transition at LHCb.  

\acknowledgments
We give our warm thanks to Aida El-Khadra for useful remarks 
on the present status of various QCD-lattice predictions.
The work of E.G. was supported in part by the U.S.\ National Science
Foundation under Grant PHY--0555304, J.H. was supported by the U.S.
Department of Energy under Contract DE-AC02-76SF00515, 
S.P. was supported by the U.S.\ Department of 
Energy under Contract DE-FG02-04ER41291 and 
A.A.P. and G.K.Y.~were supported in part by the U.S.\ National Science Foundation under
CAREER Award PHY--0547794, and by the U.S.\ Department of Energy 
under Contract DE-FG02-96ER41005.

\appendix
\section{Choice of the basis and mixing matrix elements}

There are eight $\Delta b = 2$ effective operators that can contribute 
to $B_s$-mixing.  The operator basis we shall employ is 
\beqa
\begin{array}{l}
Q_1 = (\overline{b}_L \gamma_\mu s_L) \ (\overline{b}_L \gamma^\mu s_L)\ , \\
Q_2 = (\overline{b}_L \gamma_\mu s_L) \ (\overline{b}_R \gamma^\mu s_R)\ , \\
Q_3 = (\overline{b}_L s_R) \ (\overline{b}_R s_L) \ , \\
Q_4 = (\overline{b}_R s_L) \ (\overline{b}_R s_L) \ ,
\end{array}
\qquad
\begin{array}{l}
Q_5 = (\overline{b}_R \sigma_{\mu\nu} s_L) 
\ ( \overline{b}_R \sigma^{\mu\nu} s_L)\ , \\
Q_6 = (\overline{b}_R \gamma_\mu s_R) \ (\overline{b}_R \gamma^\mu
s_R)\ , \\
Q_7 = (\overline{b}_L s_R) \ (\overline{b}_L s_R) \ , \\
Q_8 = (\overline{b}_L \sigma_{\mu\nu} s_R) \ (\overline{b}_L
\sigma^{\mu\nu} s_R)\ \ ,
\end{array}
\label{SetOfOperators}
\eeqa
where quantities enclosed in parentheses are color singlets, 
{\it e.g.} $(\overline{b}_L \gamma_\mu s_L) \equiv 
\overline{b}_{L,i} \gamma_\mu s_{L,i}$.
These operators are generated at a scale $M$ where the NP is integrated out. 
A non-trivial operator mixing then occurs via renormalization group 
running of these operators between the heavy scale $M$ and the light 
scale $\mu$ at which hadronic matrix elements are computed.

We need to evaluate the $B_s^0$-to-${\overline B}_s^0$ matrix 
elements of these eight dimension-six basis operators. This 
introduces eight {\it non-perturbative} B-parameters $\{B_i\}$ 
that require evaluation by means of QCD sum rules or QCD-lattice 
simulation.  We express these in the form 
\begin{eqnarray}\label{ME}
& & \begin{array}{l}
\langle Q_1 \rangle = {2 \over 3} f_{{\rm B}_s}^2 M_{{\rm B}_s}^2 B_1\ , \\
\langle Q_2 \rangle = - {5 \over 6} f_{{\rm B}_s}^2 M_{{\rm B}_s}^2 B_2 \ , \\
\langle Q_3 \rangle = {7 \over 12} f_{{\rm B}_s}^2 M_{{\rm B}_s}^2 B_3 \ ,\\
\langle Q_4 \rangle = - {5 \over 12} f_{{\rm B}_s}^2 M_{{\rm B}_s}^2 B_4 \ ,
\end{array}
\quad \qquad
\begin{array}{l}
\langle Q_5 \rangle = f_{{\rm B}_s}^2 M_{{\rm B}_s}^2 B_5 \ ,\\
\langle Q_6 \rangle = {2 \over 3} f_{{\rm B}_s}^2 M_{{\rm B}_s}^2 B_6 \ ,\\
\langle Q_7 \rangle = - {5 \over 12} f_{{\rm B}_s}^2 M_{{\rm B}_s}^2 B_7 \ ,\\
\langle Q_8 \rangle = f_{{\rm B}_s}^2 M_{{\rm B}_s}^2 B_8 \ \ ,
\end{array}
\end{eqnarray}
where $f_{B_s}$ is the $B_s$ meson decay constant and 
$\langle Q_i \rangle \equiv \langle {\bar B}^0_s | Q_i | 
B^0_s \rangle$. 

Ref.~\cite{Becirevic:2001xt} has performed 
a QCD-lattice determination (quenched approximation) 
of the B-parameters in an operator basis $\{ O_i \}$ which is 
distinct from the $\{ Q_i \}$ of Eq.~(\ref{SetOfOperators}),  
\begin{eqnarray}\label{SetOfOperators1}
\begin{array}{l}
O_1 = \overline{b}^i \gamma_\mu (1+\gamma_5) 
s^i \ \overline{b}^j \gamma^\mu (1+\gamma_5) s^j\ , \\
O_2 = \overline{b}^i (1+\gamma_5) s^i \ \overline{b}^j (1+\gamma_5) s^j\ , 
\\
O_3 = \overline{b}^i (1+\gamma_5) s^j \ \overline{b}^j 
(1+\gamma_5) s^i\ , 
\end{array}
\qquad 
\begin{array}{l}
\hspace {1.0cm} \phantom{x} \\
O_4 = \overline{b}^i (1+\gamma_5) s^i \ \overline{b}^j (1-\gamma_5) s^j\ , 
\\
O_5 = \overline{b}^i (1+\gamma_5) s^j \ \overline{b}^j (1-\gamma_5) s^i \ .
\end{array}
\end{eqnarray}
Three more operators $O_i\ (i=6,7,8)$ can be obtained by substituting 
right-handed chiral projection operators with the left-handed ones 
$O_i\ (i=1,2,3)$ in Eq.~(\ref{SetOfOperators1}). 
The $B_s^0$-to-${\overline B}_s^0$ matrix elements of these 
operators have been parameterized in Ref.~\cite{Becirevic:2001xt} as
\beqa
& & \begin{array}{l}
\langle O_1 \rangle = {8 \over 3} f_{{\rm B}_s}^2 
M_{{\rm B}_s}^2 \widetilde B_1\ , \\
\langle O_2 \rangle = - {5 \over 3}  R_s^2 
f_{{\rm B}_s}^2 M_{{\rm B}_s}^2 \widetilde B_2 \ , \\
\langle O_3 \rangle = {1 \over 3} R_s^2 
f_{{\rm B}_s}^2 M_{{\rm B}_s}^2 \widetilde B_3 \ ,
\end{array}
\qquad
\begin{array}{l}
\phantom{xx} \\
\langle O_4 \rangle = 2 R_s^2 f_{{\rm B}_s}^2 
M_{{\rm B}_s}^2 \widetilde B_4 \ ,\\
\langle O_5 \rangle = {2 \over 3} R_s^2  f_{{\rm B}_s}^2 
M_{{\rm B}_s}^2 \widetilde B_5 \ .
\label{neweq}
\end{array}
\eeqa
%
Also, the chiral structure of QCD requires that 
$\langle O_6 \rangle = \langle O_1 \rangle$, 
$\langle O_7 \rangle = \langle O_2 \rangle$, 
and $\langle O_8 \rangle = \langle O_3 \rangle$.

Several of the quantities introduced above are scale dependent, 
{\it i.e.} $\{ B_i (\mu)\}$, $\{ {\widetilde B}_i (\mu)\}$ and 
$R_s^2(\mu)$.  Throughout this paper, we shall understand all 
these quantities to be 
renormalized at a common scale $\mu = m_b$ and to simplify   
notation, we shall denote them simply as $\{ B_i \}$, 
$\{ {\widetilde B}_i \}$ and $R_s^2$.  In particular, 
our evaluation at scale $\mu = m_b$ of the quantity 
$R_s(\mu) \equiv M_{B_s}/(m_b(\mu)+m_s(\mu))$ yields 
\beq
R_s^2 = M_{B_s}^2/\left({\bar m}_b ({\bar m}_b) 
+ {\bar m}_s ({\bar m}_b)\right)^2 = 1.57_{-0.10}^{+0.04} \ \ ,
\label{rs}
\eeq
where we have used the input values ${\bar m}_b ({\bar m}_b) 
= 4.2^{+0.17}_{-0.07}$~GeV~\cite{PDG} and 
${\bar m}_s ({\bar m}_b) = 0.085 \pm 0.017$~~GeV~\cite{Lenz:2006hd}.

The two bases $\{Q_i\}$ and $\{O_i\}$ can be related via 
Fierz rearrangement,  
\begin{eqnarray}\label{OperatorRelations}
& & \begin{array}{l}
O_1 = 4 \ Q_1 \ , \\
O_2 = 4 \ Q_4 \ , \\
O_3 = - 2 \ Q_4 - \frac{1}{2} \ Q_5 \ ,
\end{array}
\qquad
\begin{array}{l}
O_4 = 4 \ Q_3 \ ,\\
O_5 = - 2 \ Q_2 \ .
\end{array}
\end{eqnarray}
from which we find 
\begin{eqnarray}\label{BagRelations}
& & \begin{array}{l}
B_1 = \widetilde B_1 \ , \\
B_2 = \frac{2}{5} \widetilde B_5 R_s^2 \ , \\
B_3 = \frac{6}{7}  \widetilde B_4 R_s^2 \ , \\
B_4 = \widetilde B_2 R_s^2 \ ,
\end{array}
\qquad
\begin{array}{l}
B_5 = - \frac{1}{3} R_s^2 \left(2 \widetilde B_3 -  
5 \widetilde B_2\right) \ , \\
B_6 = \widetilde B_1 \ , \\
B_7 = \frac{6}{7}  \widetilde B_4 R_s^2\ , \\
B_8 = - \frac{1}{3} R_s^2 \left(2 \widetilde B_3 -  
5 \widetilde B_2\right) \ .
\end{array}
\end{eqnarray}
\begin{table}[t]
\begin{tabular}{l||c|c}
\colrule\hline 
List of $\{ B_i\}$  & 
$\{ B_i\}$ from lattice QCD & $B_i$ in MVS \\ 
(in $\{ Q_i\}$ Basis) & 
(from Ref.~\cite{Becirevic:2001xt})  &  (from Eq.~(\ref{ME_MVS})) \\
\colrule \colrule
$B_1 = B_6$ & $0.87$ & $0.87$ \\
$B_2$ & $0.70 R_s^2$ & $0.87 \left[ {3 \over 5} + 
{2 \over 5} R_s^2\right]$ \\
$B_3$ & $0.99 R_s^2$ & $0.87 \left[ {1 \over 7} + 
{6 \over 7} R_s^2\right]$ \\
$B_4 = B_7$ & $0.80 R_s^2$ & $0.87 R_s^2$ \\
$B_5 = B_8$ & $0.71 R_s^2$ & $0.87 R_s^2$ \\
\colrule\hline
\end{tabular}
\vskip .05in\noindent
\caption{Numerical estimates of the B-parameters. 
The determination from lattice QCD is done in 
$\overline{\rm MS}$(NDR).} 
\label{tab:BagConstants}
\end{table}
Alternatively, the B-parameters can be estimated using the 
`modified vacuum saturation' (MVS) approach, wherein 
all matrix elements in Eq.~(\ref{ME}) are written in terms of (known)
matrix elements of $(V-A)\times (V-A)$ and $(S-P)\times (S+P)$ matrix
elements $B_{\rm B}$ and $B_{\rm B}^{\rm (S)}$, 
\begin{eqnarray}\label{ME_MVS}
& & \begin{array}{l}
\langle Q_1 \rangle = \displaystyle{2 \over 3} 
f_{{\rm B}_s}^2 M_{{\rm B}_s}^2 B_{B_s} \ ,\\
\langle Q_2 \rangle = f_{{\rm B}_s}^2 M_{{\rm B}_s}^2 B_{B_s} \left[ 
- \displaystyle{1 \over 2} - \displaystyle{\eta \over N_c} \right] \ ,\\
\langle Q_3 \rangle = f_{{\rm B}_s}^2 M_{{\rm B}_s}^2 B_{B_s} \left[ 
\displaystyle{1 \over 4 N_c} + \displaystyle{\eta \over 2} \right] \ ,\\
\langle Q_4 \rangle = - \displaystyle{2 N_c - 1 \over 4 N_c} 
f_{{\rm B}_s}^2 M_{{\rm B}_s}^2 B_{B_s}~ \eta \ ,
\end{array}
\qquad \qquad
\begin{array}{l}
\langle Q_5 \rangle = \displaystyle{3 \over N_c} 
f_{{\rm B}_s}^2 M_{{\rm B}_s}^2 B_{B_s} ~ \eta \ , \\
\langle Q_6 \rangle = \langle Q_1 \rangle \ , \\
\langle Q_7 \rangle = \langle Q_4 \rangle \ , \\
\langle Q_8 \rangle = \langle Q_5 \rangle \ \ ,
\end{array}
\end{eqnarray}
where we take $N_c=3$ as the number of colors and define 
\beqa\label{bbar}
& & \eta \equiv {B_{{\rm B}_s}^{\rm (S)}\over B_{B_s}} 
 \cdot {M_{{\rm B}_s}^2 \over \left({\bar m}_b ({\bar m}_b) 
+ {\bar m}_s ({\bar m}_b)\right)^2} 
\to R_s^2  \ \ \text{for} \ \ B_{{\rm B}_s}^{\rm (S)} = B_{B_s}
 \ \ . 
\eeqa
It is instructive to compare how well the MVS approximation 
estimates the recent lattice results. We provide such 
a comparison in Table~\ref{tab:BagConstants}.


\end{document}